\documentclass[leqno,10pt]{crmp-l} 
\usepackage{amscd,amssymb,amsthm,amsmath,epsfig}

\newcommand{\ovl}{\overline}
\newcommand{\n}{\noindent}
\newcommand{\ds}{\displaystyle}

\newcommand{\bb}[1]{\mathbb{#1}}
\newcommand{\cl}[1]{\mathcal{#1}}
\newcommand{\brel}[1]{\pmb{#1}}

\theoremstyle{plain}
\newtheorem{thm}{Theorem}[section]
\newtheorem{lem}[thm]{Lemma}
\newtheorem{cor}[thm]{Corollary}

\theoremstyle{definition}
\newtheorem{defn}{Definition}[section]
\newtheorem{exm}{Example}[section]

\theoremstyle{remark}
\newtheorem{rem}{Remark}[section]

\numberwithin{equation}{section}
\numberwithin{figure}{section}

\begin{document}

\title[Mathematical Models]{Mathematical Models of Contemporary Elementary 
Quantum Computing Devices}

\author{G. Chen}
\address{Department of Mathematics, Texas A\&M University, College Station, 
TX~77843, U.S.A.}
\email{gchen@math.tamu.edu}
\author{D. A. Church}
\address{Department of Physics, Texas A\&M University, College Station, 
TX~77843, U.S.A.}
\email{church@physics.tamu.edu}
\author{B.-G. Englert}
\address{Department of Physics, National University of Singapore, Singapore}
\email{phyebg@nus.edu.sg}
\author{M. S. Zubairy}
\address{Department of Physics, Texas A\&M University, College Station,
 TX~77843, U.S.A.} 
\email{zubairy@physics.tamu.edu}
\thanks{The first and fourth authors are also of the Institute of Quantum 
Studies, Texas A\&M University, College 
Station, TX 77843, U.S.A., and are supported in part by DARPA QuIST Contract
F49620-01-1-0566 and Texas A\&M University TITF initiative.
The third author was supported in part by NSF Grant PHY9876899.}

\begin{abstract}
Computations with a future quantum computer will be implemented through 
the operations by elementary quantum gates. It is now well known that the 
collection of 1-bit and 2-bit quantum gates are universal for 
quantum computation, i.e., any $n$-bit unitary operation can be carried out 
by concatenations of 1-bit and 2-bit elementary quantum gates.

Three contemporary quantum devices--cavity QED, ion traps and quantum 
dots--have been widely regarded as perhaps the most promising candidates 
for the construction of elementary quantum gates. In this paper, we describe 
the physical properties of these devices, and show the mathematical 
derivations based on the interaction of the laser field as control  with 
atoms, ions or electron spins, leading to the following:
\begin{itemize}
\item[(i)] the 1-bit unitary rotation gates; and 
\item[(ii)] the 2-bit quantum phase gates and the controlled-not gate.
\end{itemize}
This paper is aimed at providing a sufficiently self-contained survey 
account of analytical nature for mathematicians, physicists and computer 
scientists to aid interdisciplinary understanding in the 
research of quantum computation.
\end{abstract}

\maketitle

\section{Introduction}\label{sec1}

The design and construction of the quantum computer is a major project by 
the scientific community of the 21$^{\rm st}$ Century. This project is 
interdisciplinary by nature, requiring close collaboration between 
physicists, engineers, computer scientists and mathematicians.

Physicists have taken the lead in the design of quantum devices utilizing the 
peculiar properties of quantum mechanics. Most or perhaps all of such 
devices will eventually become micro- or nano-fabricated. Their controls are 
achieved through the 
tuning of laser pulses. As hardware, many types of material or 
devices have been either used or proposed: NMR, cavity QED, ion or atom 
traps, quantum dots, SQUID, etc. See some discussions in \cite[Chap.~7]{NC}, 
\cite{S}, for example. NMR (nuclear magnetic resonance) is the first
scheme utilized by researchers to demonstrate the principle of quantum
computing through quantum search.  
NMR seems to have provided the most successful demonstration of quantum 
computing so far.
However, this kind of 
``quantum computing in a coffee mug'' utilizes bulk material, which is not 
considered as really quantum by many people. In addition, NMR has a severe 
weakness in the lack of scalability. 
Concerning SQUIDs\footnote{The SQUID can be 
viewed as a high-$Q$
microwave frequency LC resonator.  
The first excited state, with excitation energy $\hbar \omega$
where $\omega$ is the microwave frequency, is doubly degenerate in zero
magnetic field.  These two states are the qubit states. To lift 
the degeneracy, Josephson junction(s) are placed in the current loop.  
Magnetic field can then be used to tune the levels back to degeneracy 
as required for qubit coupling, etc.  
The coupling is magnetic, as in mutual inductance of two current loops.
 
The major breakthrough made last year (2001--02) is the single quantum 
flux gate for readout.  
It is 100\% efficient and does not add decoherence.  Expect to 
see rapid progress in the near future as a result of this breakthrough.  

We thank Prof. Philip R. Hemmer for the above communication.} 
(superconducting quantum interference devices), there 
has been major progress in the design and control of such devices 
recently. 
However,  the authors' analytical knowledge about SQUIDs still appears 
very limited for the time being and, thus, any detailed mathematical 
representations must be deferred to a future account.

We focus our attention on the following three quantum 
devices: cavity QED, ion traps and quantum dots. They seem to have 
received the most attention in the contemporary literature, and have been
widely identified by many as the most promising. The writing of this paper 
all began with our 
simple desire to try to understand, analytically, how and why 
such devices work.  Analytical studies of these devices are scattered in 
many references in the physics literature, which admittedly are not easy 
reading for most  mathematicians. Indeed, this is the most commonly 
encountered difficulty in any interdisciplinary research. But even 
many physicists specializing in quantum devices find certain difficulty 
in their attempt to understand the working of other quantum devices, if such 
devices do not fall exactly within their own specialty. Therefore, our 
objectives in this paper are three-fold:
\begin{itemize}
\item[(1)] Provide a sufficiently self-contained account of the physical 
description of these primary contemporary quantum computing devices and 
the derivation of their mathematical representations. 
\item[(2)] Supply the control-theoretic aspect of quantum control via the 
shaping 
of laser pulses in quantum computing, which is the main theme of the 
conference.
\item[(3)] Write an accessible survey of three important quantum devices for 
mathematicians and other scientists who are interested in understanding 
the basic interdisciplinary link between physics, mathematics and computer 
science, in a mathematically rigorous way, for quantum computing.
\end{itemize}

Our main emphasis will be 
on the mathematical modeling. As far as the physics is concerned 
in this paper, we acknowledge the many segments 
that we have extracted, directly or indirectly, from the original sources 
cited in the references. New devices and designs emerge almost daily, and we 
must concede that the list of 
references is neither exhaustively comprehensive nor totally up-to-date as 
any attempt to achieve such by us is nearly impractical, or even 
impossible given the time and various other constraints. We 
nevertheless believe that the basic mathematics underlying most of the quantum 
computing devices remains largely similar, and hope that we have provided 
enough ideas and information for the readers 
to trace the literature on their own.

In the main body of this paper to follow, there are four sections. 
Section 2 provides a quick summary of universality results of 
quantum gates. Sections 3, 4 and 5 deal with, respectively, cavity QED, ion 
traps and quantum dots. In the final Section 6, we briefly discuss some 
issues on laser control in quantum computing as the conclusion of this 
paper.

\section{Universality of Elementary Quantum Gates}\label{sec2}

We begin this section by introducing some basic notations used in quantum 
mechanics, as such notations do not appear to be familiar to a majority of 
mathematicians. A useful reference can be found in \cite{SZ}.

Let $\cl H$ be a {\em complex\/} Hilbert space with inner product $\langle 
~~,~~\rangle$. For any subset $S\subseteq \cl H$,  define
\begin{align}
\text{span } S &= \text{closure of } \left\{\sum^n_{j=1} \alpha_j\psi_j \mid 
\alpha_j\in\bb C, \psi_j \in S, j=1,2,\ldots, n, \text{ for all } \psi_j\in 
S\right\}\nonumber\\
\label{eq2.1}
&\qquad \text{ in } \cl H,
\end{align}
where in the above, $\bb C$ denotes the set of all complex numbers. 
Elements in 
$\cl H$, such as $\psi_j$ in \eqref{eq2.1}, are called vectors. The inner 
product of two vectors $\psi_j$ and $\psi_k$ is $\langle\psi_j,\psi_k\rangle$. 
However, in quantum mechanics, the Dirac bra-ket notation writes the vector 
$\psi_k$ as $|k\rangle$, pronounced {\em ket\/} $k$, where $k$ labels the 
vector. The role of 
$\psi_j$ in the inner product $\langle\psi_j,\psi_k\rangle$ is 
that $\psi_j$ is identified as a linear functional on $\cl H$ by
\begin{equation}
\psi^*_j\colon \ \cl H\longrightarrow \bb C;\quad \psi^*_j(\psi_k) \equiv 
\langle\psi_j,\psi_k\rangle.
\end{equation}
The Dirac bra-ket  notation writes $\psi^*_j$ as $\langle j|$, pronounced 
{\em bra\/} $j$. The inner product is now written as
\begin{equation}
\langle\psi_j,\psi_k\rangle = \langle j|k\rangle.
\end{equation}
Let $A\colon \ \cl H\to \cl H$ be a linear operator. Then in Dirac's notation 
we 
write $\langle \psi_j, A\psi_k\rangle$ as $\langle j|A|k\rangle$.

The set of all kets $(|k\rangle)$ generates the linear space $\cl H$, and the 
set of all bras $(\langle j|)$ also generates the dual linear space $\cl H^*$. 
These two linear spaces are called, respectively, the ket space and the bra 
space, which are dual to each other. The natural isomorphism from $H$ onto 
$H^*$, or from $H^*$ onto $H$, is called ``taking the adjoint'' by physicists. 
An expression like $\langle j|A|k\rangle$ has an intended symmetry:\ one can 
think of $A$ acting on $|k\rangle$ on the right as a matrix times a column 
vector, or $A$ acting on $\langle j|$ on the left as a row vector times  a 
matrix.

Any  2-dimensional complex Hilbert space $\cl H$ is isomorphic to $\bb C^2$. 
The standard basis of $\bb C^2$ is
\begin{equation}
\pmb{e}_1 = \left[\begin{matrix} 1\\ 0\end{matrix}\right],\quad \pmb{e}_2 = 
\left[\begin{matrix} 0\\ 1\end{matrix}\right].
\end{equation}
In quantum mechanics, the preferred notation for basis is, respectively, 
$|0\rangle$ and 
$|1\rangle$, 
where $|0\rangle$ and $|1\rangle$ generally refer to the ``spin'' state of a 
certain quantum system under discussion.
``Spin up'' and ``spin down'' 
form a binary quantum alternative and, therefore, a quantum bit or, 
a \emph{qubit}. These spin states may then further be 
identified with $\pmb{e}_1$ and $\pmb{e}_2$ through, e.g.,
\begin{equation}\label{eq2.1a}
|0\rangle = \left[\begin{matrix} 1\\ 0\end{matrix}\right],\quad |1\rangle = 
\left[\begin{matrix} 0\\ 1\end{matrix}\right].
\end{equation}
For a column vector, say $\pmb{e}_1$, 
we use superscript $T$ to denote its transpose, 
such as $\pmb{e}^T_1 = [1~~0]$, 
for example. The spin states $|0\rangle$ and $|1\rangle$ form a standard 
orthonormal basis for the (single) qubit's ket space. The tensor product of a 
set of $n$ vectors 
$|z_j\rangle\in 
\bb C^2_j$, specified by the quantum numbers $z_j$
for $j=1,2,\ldots, n$, is written interchangeably as
\begin{equation}\label{eq2.2}
|z_1\rangle \otimes |z_2\rangle \otimes\cdots \otimes |z_n\rangle = 
|z_1\rangle |z_2\rangle \cdots |z_n\rangle = |z_1z_2 \cdots 
z_n\rangle.
\end{equation}
The tensor product space of $n$ copies of $\bb C_2$ is defined to be
\begin{equation}\label{eq2.3}
(\bb C_2)^{\otimes n} = \text{span}\{|j_1j_2\cdots j_n\rangle \mid j_k\in 
\{0,1\}, k=1,2,\ldots, n\}.
\end{equation}
It is a $2^n$-dimensional complex Hilbert space with the induced inner product 
from $\bb C^2$. A {\em quantum state\/} is a vector in $(\bb C^2)^{\otimes n}$ 
(or in any Hilbert space $\cl H$) with the unit norm\footnote{This type of 
quantum state is actually a more restrictive type, called a 
\emph{pure state}. 
A pure state, say ket $|k\rangle$, could be identified with the projection 
operator $|k\rangle\langle k|$. 
Then a \emph{general state} is a convex sum of 
such projectors, i.e., $\sum\limits_k c_k|k\rangle\langle k|$ with $c_k\ge 0$ 
and $\sum\limits_k c_k=1$.}. 
A quantum state is said to be \emph{entangled} if it is \emph{not} 
a tensor product of the form \eqref{eq2.2}. 
A \emph{quantum operation} is a \emph{unitary linear transformation} 
on $\cl H$. 
The \emph{unitary group} $U(\cl H)$ on $\cl H$ consists of all unitary linear 
transformations on $\cl H$ with composition as the (noncommutative) natural 
multiplication operation of the group.

It is often useful to write the basis vectors $|j_1j_2\cdots j_n\rangle \in 
(\bb 
C^2)^{\otimes n}$, $j_k\in \{0,1\}$, $k=1,2,\ldots, n$, as column vectors 
according to the lexicographic ordering:
\begin{align}
|00\cdots 00\rangle &= \left[\begin{matrix} 1\\ 0\\ 0\\ \vdots\\ 0\\ 0\\ 
0\end{matrix}\right] , |00\cdots 01\rangle = \left[\begin{matrix} 0\\ 1\\ 0\\ 
\vdots\\ 0\\ 0\\ 0\end{matrix}\right], |00\cdots 10\rangle = 
\left[\begin{matrix} 0\\ 0\\ 1\\ \vdots\\ 0\\ 0 \\ 0\end{matrix}\right], 
\cdots\nonumber\\
\label{eq2.4}
|11\cdots 10\rangle &= \left[\begin{matrix} 0\\ 0\\ 0\\ \vdots\\ 0\\ 1\\ 
0\end{matrix}\right] , |11\cdots 11\rangle = \left[\begin{matrix} 0\\ 0\\ 0\\ 
\vdots\\ 0\\ 0\\ 1\end{matrix}\right];
\end{align}
cf.\ \eqref{eq2.1a}. Any $2\times 2$ unitary matrix is an admissible quantum 
operation on one qubit. 
We call it a 1-bit gate. Similarly, we call a $2^k\times 2^k$  unitary 
transformation a $k$-bit gate. Let a 1-bit gate be
\begin{equation}
U = \left[\begin{matrix} u_{00}&u_{01}\\ u_{10}&u_{11}\end{matrix}\right];
\end{equation}
we define the operator $\Lambda_m(U)$ \cite{Ba} on $(m+1)$-qubits (with 
$m=0,1,2,\ldots$) through its action on the basis by
\begin{equation}
\Lambda_m(U)(|x_1x_2\cdots x_my\rangle) = 
\left\{\begin{array}{@{}l@{\ }l@{\ }l@{}}
|x_1x_2\cdots x_my\rangle&\text{if}&\wedge^m_{k=1} x_k=0,\\
u_{0y}|x_1x_2\cdots x_m0\rangle + u_{1y}|x_1x_2\cdots x_m1\rangle
&\text{if}&
\wedge^m_{k=1}x_k=1,\end{array}\right.
\end{equation}
where ``$\wedge$'' denotes the Boolean operator AND. The matrix representation 
for $\Lambda_m(U)$ according to the ordered basis \eqref{eq2.4} is
\begin{equation}
\Lambda_m(U) = \left[\begin{matrix}
1\\ &\ddots&&\bigcirc\\ &&1\\ &\bigcirc&&\left[\begin{matrix} u_{00}&u_{01}\\ 
u_{10}&u_{11}\end{matrix}\right]\end{matrix}\right]_{2^{m+1}\times 2^{m+1}}.
\end{equation}
This unitary operator is called the $m$-bit-controlled $U$ operation.

An important logic operation on one qubit is the NOT-gate
\begin{equation}
\sigma_x = \left[\begin{matrix} 0&1\\ 1&0\end{matrix}\right];
\end{equation}
cf.\ the Pauli matrices in \eqref{eq5.1a} of Section \ref{sec5}. From 
$\sigma_x$, 
 define the important 2-bit operation $\Lambda_1(\sigma_x)$, which is the 
controlled-not gate, henceforth acronymed the CNOT gate.

For any 1-bit gate $A$,  denote by $A(j)$ the operation (defined through
its action on the basis) on the slot $j$:
\begin{equation}
A(j)|x_1x_2\cdots x_j\cdots x_n\rangle = |x_1\rangle \otimes |x_2\rangle 
\otimes\cdots \otimes |x_{j-1}\rangle \otimes [A|x_j\rangle] \otimes |x_{j+1} 
\rangle \otimes\cdots\otimes |x_n\rangle.
\end{equation}
Similarly, for a 2-bit gate $B$, we define $B(j,k)$ to be the operation on the 
two qubit slots $j$ and $k$ \cite{BB}. A simple 2-bit gate is the swapping gate
\begin{equation}\label{eq2.5}
U_\mathrm{sw}
|x_1x_2\rangle = |x_2x_1\rangle, \text{ for } x_1,x_2\in \{0,1\}.
\end{equation}
Then $U_\mathrm{sw}(j,k)$ swaps the qubits between the $j$-th 
and the $k$-th slots.

The tensor product of two 1-bit quantum gates $S$ and $T$ is defined through
\begin{equation}
(S\otimes T) (|x\rangle \otimes |y\rangle = (S|x\rangle) \otimes (T|y\rangle), 
\text{ for } x,y\in \{0,1\}.
\end{equation}
A 2-bit quantum gate $V$ is said to be \emph{primitive} 
(\cite[Theorem 4.2]{BB}) if
\begin{equation}
V = S\otimes T\quad \text{or}\quad V = (S\otimes T) U_\mathrm{sw}
\end{equation}
for some 1-bit gates $S$ and $T$. Otherwise, $V$ is said to be 
\emph{imprimitive}.

It is possible to factor an $n$-bit quantum gate $U$ as the composition of 
$k$-bit 
quantum gates for $k\le n$. The earliest result of this {\em universality\/} 
study was given by Deutsch \cite{De} in 1989 for $k=3$. Then in 1995, Barenco 
\cite{BaO}, Deutsch, Barenco and Ekert \cite{DBE}, DiVincenzo \cite{DiV} and 
Lloyd \cite{Ll} gave the result for $k=2$. Here, we quote the elegant 
mathematical treatment and results by J.-L. Brylinski and R.~K. Brylinski in 
\cite{BB}.

\begin{defn}[$\lbrack$\ref{BB}, Def.~4.2$\rbrack$]\label{defn2.1}
A collection of 1-bit gates $A_i$ and 2-bit gates $B_j$ is called (exactly) 
{\em universal\/} if, for each $n\ge 2$, any $n$-bit gate can be obtained by 
composition of the gates $A_i(\ell)$ and $B_j(\ell,m)$, for 
$1\le \ell, m\le n$. \hfill$\square$
\end{defn}

\begin{thm}[$\lbrack$\ref{BB}, Theorem 4.1$\rbrack$]\label{thm2.1}
Let $V$ be a given 2-bit gate. Then the following are equivalent:
\begin{itemize}
\item[(i)] The collection of all 1-bit gates $A$ together with $V$ is 
universal.
\item[(ii)] $V$ is imprimitive.$\hfill\square$
\end{itemize}
\end{thm}

A common type of 1-bit gate from AMO (atomic, molecular and optical) 
devices is the unitary rotation gate
\begin{equation}\label{eq2.6}
U_{\theta,\phi} \equiv \left[\begin{matrix} \cos\theta&-ie^{-i\phi}\sin\theta\\
-ie^{i\phi}\sin\theta&\cos\theta\end{matrix}\right],
\qquad 0\le \theta,\phi \le 2\pi.
\end{equation}
We have (the determinant) $\det U_{\theta,\phi} = 1$ for any $\theta$ and 
$\phi$ 
and, thus, the collection of all such $U_{\theta,\phi}$ is not dense in $U(\bb 
C^2)$. Any unitary matrix with determinant equal to 1 is said to be {\em 
special 
unitary\/} and the collection of all special unitary matrices on $(\bb 
C^2)^{\otimes n}$, $SU((\bb C^2)^{\otimes n})$, 
is a proper subgroup of $U((\bb 
C^2)^{\otimes n})$. It is known that the gates $U_{\theta,\phi}$ generate 
$SU(\bb C^2)$.

Another common type of 2-bit gate we shall encounter in the next two sections
is the quantum phase gate (QPG) 
\begin{equation}\label{eq2.7}
Q_\eta = \left[\begin{matrix} 1&0&0&0\\ 0&1&0&0\\ 0&0&1&0\\ 
0&0&0&e^{i\eta}\end{matrix}\right],\qquad 0\le \eta\le 2\pi.
\end{equation}

\begin{thm}[$\lbrack$\ref{BB}, Theorem 4.4$\rbrack$]\label{thm2.2}
The collection of all the 1-bit gates $U_{\theta,\phi}$, $0\le \theta,\phi$, 
$\phi\le 2\pi$, together with any 2-bit gate $Q_\eta$ where $\eta\not\equiv 0$ 
(mod $2\pi$), is universal.$\hfill\square$
\end{thm}

Note that the CNOT gate $\Lambda_1(\sigma_x)$ can be written as
\begin{equation}\label{eq2.8}
\Lambda_1(\sigma_x) = U_{\pi/4,\pi/2}(2) Q_\pi U_{\pi/4,-\pi/2}(2);
\end{equation}
cf.\ \cite[(2.2)]{DZC}. Thus, we also have the following.

\begin{cor}\label{cor2.3}
The collection of all the 1-bit gates $U_{\theta,\phi}$, $0\le \theta,\phi \le 
2\pi$, together with the CNOT gate $\Lambda_1(\sigma_x)$, is universal. 
$\hfill\square$
\end{cor}

\begin{figure}[ht]
\centerline{\epsfig{file=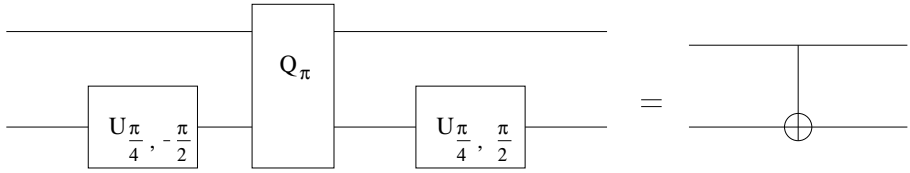,width=0.95\textwidth}}  
\caption{\label{fig2.1}%
The quantum circuits for equation \eqref{eq2.8}. 
The top wire represents the leading qubit (which is the most significant), 
while the second wire represents the second qubit.}
\end{figure}

Diagramatically, the quantum circuit for \eqref{eq2.8} is given in 
Fig.~\ref{fig2.1}.

\section{Two-Level Atoms and Cavity QED}\label{sec3}

The main objective of this section is to show that the 1-bit unitary gates 
\eqref{eq2.6}  and the 2-bit QPG \eqref{eq2.7} can be implemented using the 
methods of, respectively, 2-level atoms and cavity QED. For clarity, we divide 
the discussions in three subsections.

\subsection{Two-Level Atoms}\label{sec3.1}

The atomic energy levels are very susceptible 
to excitation by electromagnetic radiation. The 
structure of electronic eigenstates and the interaction between electrons and 
photons can be quite complicated. But under certain assumptions we have in 
effect a 2-level atom. These assumptions are as follows:
\begin{itemize}
\item[(i)] the difference in energy levels of the  electrons matches the 
energy of the incident photon; 
\item[(ii)] the symmetries of the atomic structure and the resulting 
``selection rules'' allow the transition of the electrons between the two 
levels; and 
\item[(iii)] all the other levels are sufficiently ``detuned'' in frequency 
separation with respect to 
the frequency of the incident field such that there is no transition to those 
levels.
\end{itemize}
Then the model of a  2-level atom provides a good approximation. 

\begin{figure}[ht]
\centerline{\epsfig{file=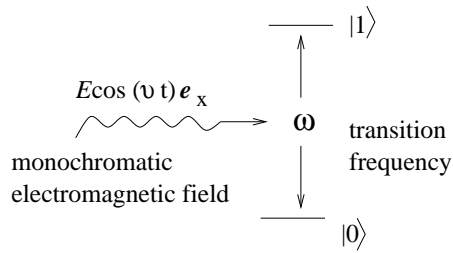,height=1.3in}}
\caption{\label{fig3.1}%
A 2-level atom, where a bound electron  in an atom
interacts with an external electromagnetic field.}
\end{figure}

The wave function $\psi(\pmb{r},t)$ of the electron of the 2-level atom,
which is coupled to the external electric field $\pmb{E}(\pmb{r},t)$ by its
electric charge $e$, as shown in Fig.~\ref{fig3.1}, obeys the 
Schr\"odinger equation
\begin{equation}\label{eq3.3}
i\hbar \frac\partial{\partial t} \psi(\pmb{r},t) = H\psi(\pmb{r},t) 
= (H_0+H_1) \psi(\pmb{r},t),
\end{equation}
where $|\psi(\pmb{r},t)|^2$ is the probability density 
of finding that electron at position $\pmb{r} = (x,y,z)$ at time $t$, and
\begin{align}\label{eq3.6c}
H_0 &\equiv \frac1{2m} \pmb{\nabla}^2 + V(\pmb{r}) &
\begin{minipage}[t]{200pt}
is the differential operator for the unperturbed Hamiltonian of the electron,
with the gradient operator 
$\pmb{\nabla}=(\partial/\partial x, \partial/\partial y, \partial/\partial
z)$
and the electrostatic potential $V(\pmb{r})$ between 
the electron and the nucleus,
\end{minipage}
\\
\label{eq3.6d}
H_1 &\equiv -e\pmb{r} \cdot \pmb{E}(\pmb{r}_0,t) & 
\begin{minipage}[t]{200pt}
is the interaction 
Hamiltonian between the field and the electron of the atom.
\end{minipage}
\end{align}

\begin{rem}\label{rem3.0.c}
The wavelength of visible light, typical for atomic transitions, 
is about a few 
thousand times the diameter of an atom.  Therefore, there is no significant 
spatial variation of the electric field across an atom and so we can replace 
$\pmb{E}(\pmb{r},t)$ by $\pmb{E}(\pmb{r}_0,t)$, the field at a reference point 
inside the atom such as the position of the nucleus or the center of mass. 
Consistent with this long-wavelength approximation (known as {\em dipole 
approximation\/} in the physics literature) is that the magnetic field 
$\pmb{B}$ satisfies $\pmb{B}(\pmb{r},t) \cong 0$, so that the Lorentz force of 
the radiation field on the electron is $e\pmb{E}(\pmb{r}_0,t)$, the negative 
gradient of $-e\pmb{r}\cdot \pmb{E}(\pmb{r}_0,t)$, which adds to the potential 
energy $V(\pmb{r})$ in the Hamiltonian operator.$\hfill\square$
\end{rem}

Let the external electromagnetic field be a monochromatic plane-wave field 
linearly polarized along the $x$-axis, which interacts with the atom placed at 
$\pmb{r}_0 = \pmb{0}$. The electric field takes the form 
\begin{equation}\label{eq3.6e}
\pmb{E}(\pmb{0},t) 
= \cl E\cos(\nu t) \pmb{e}_x;\quad \pmb{e}_x \equiv [1,0,0]^T.
\end{equation}
Let $|1\rangle$ and $|0\rangle$ represent upper and lower level states of the 
atom, which are eigenstates of the unperturbed part $H_0$ of the Hamiltonian 
corresponding to the eigenvalues $\hbar \omega_1$ and $\hbar \omega_0$, 
respectively, with $\omega \equiv \omega_1-\omega_0$, cf.\ Fig.~\ref{fig3.1}.

For the unperturbed 2-level atom system, i.e., without the external 
electromagnetic field, 
let the eigenstates of the electron be $|\psi_0\rangle = 
|0\rangle$ and $|\psi_1\rangle= |1\rangle$, i.e.,
\begin{equation}\label{eq3.7}
\left\{\begin{array}{lll}
H_0|\psi_0\rangle = E_0|\psi_0\rangle,&\text{or}&H_0|0\rangle = \hbar 
\omega_0|0\rangle;\qquad 
\langle 0|0\rangle = 1,\\[1ex]
H_0|\psi_1\rangle = E_1|\psi_1\rangle,&\text{or}&H_0|\psi_1\rangle = \hbar 
\omega_1|1\rangle; 
\qquad \langle 1|1\rangle = 1.\end{array}\right.
\end{equation}
We invoke the assumption of 2-level atom that the electron 
lives only on states 
$|0\rangle$ and $|1\rangle$:
\begin{equation}\label{eq3.8}
|\psi(t)\rangle  = C_0(t)|0\rangle + C_1(t)|1\rangle,
\end{equation}
where $C_0(t)$ and $C_1(t)$ are complex-valued functions such that $|C_0(t)|^2 
+ 
|C_1(t)|^2 = 1$, with $|C_0(t)|^2$ and $|C_1(t)|^2$ being the probability of 
the 
electron in, respectively, state $|0\rangle$ and $|1\rangle$ at time $t$.

Since $H_1$ (after quantization) is an operator effective only on the subspace
\begin{equation}
S = \text{span}\{|0\rangle, |1\rangle\},
\end{equation}
$H_1$ is the zero operator on $S^\bot$ (the orthogonal complement of $S$ 
in the Hilbert space $\cl H$ which 
has an orthonormal basis consisting of all eigenstates of $H_0$).  We utilize 
the property that on $S$ the projection operator is 
$|0\rangle \langle 0| + |1\rangle\langle 1|$, and obtain 
\begin{align}
H_1 &= - e\pmb{r}\cdot \pmb{E}(\pmb{0},t) = -e(x\pmb{e}_x + y\pmb{e}_y + 
z\pmb{e}_z)\cdot (\cl E\cos (\nu t)\pmb{e}_x) \nonumber\\
&= -e\cl Ex \cos(\nu t) \text{ (by \eqref{eq3.6d} and 
\eqref{eq3.6e})}\nonumber\\
&= -e\cl E[|0\rangle\langle 0| + |1\rangle\langle 1|] x [|0\rangle \langle 
0| + |1\rangle\langle 1|] \cos (\nu t)\nonumber\\
&= -e\cl E\{[\langle 0| x |0\rangle] |0\rangle\langle 0| + [\langle 0| x 
|1\rangle] |0\rangle \langle 1|\nonumber\\
\label{eq3.9}
&\quad + [\langle 1| x |0\rangle] |1\rangle\langle 0| + [\langle 1| x
|1\rangle] |1\rangle\langle 1|\} \cos (\nu t).
\end{align}
But from the symmetry property and the selection rules that 
$|\psi_j(\pmb{r})|^2 = |\psi_j(-\pmb{r})|^2$ for $j=0$ 
and 1, we have
\begin{align}
\langle 0| x |0\rangle = \int\limits^\infty_{-\infty}
\int\limits^\infty_{-\infty}\int\limits^\infty_{-\infty} x
|\psi_0(\pmb{r})|^2 dxdydz =0
\end{align}
since  $x=\pmb{r}\cdot\pmb{e}_x$ is an odd function of $\pmb{r}$ whereas
$|\psi_0(\pmb{r})|^2$ is even.
Similarly, $\langle 1|x |1\rangle = 0$. Now, write $P_{01} \equiv e\langle 
0| x |1\rangle$ and $P_{10} \equiv e\langle 1|x |0\rangle$. Then 
$P_{10} = \ovl P_{01}$. To solve the partial differential
differential equation \eqref{eq3.3}, we need only focus our attention on the
2-dimensional invariant subspace $S$ and, thus, a dramatic reduction of 
dimensionality. Substituting \eqref{eq3.8} into \eqref{eq3.3}, utilizing 
\eqref{eq3.7} and \eqref{eq3.9} and simplifying, we obtain a $2\times 2$ first 
order linear ODE
\begin{equation}\label{eq3.10}
\frac{d}{dt} {\left[\begin{matrix} C_0(t)\\ C_1(t)\end{matrix}\right]} = 
\left[\begin{matrix} -i\omega_0&i\Omega_R e^{-i\phi}\cos \nu t\\ i\Omega_R 
e^{i\phi} \cos \nu t&-i\omega_1\end{matrix}\right] \left[\begin{matrix} 
C_0(t)\\ C_1(t)\end{matrix}\right],
\end{equation}
where
\begin{equation}\label{eq3.11}
\Omega_R \equiv \frac{|P_{10}| \cl E}\hbar  = \text{the Rabi frequency},
\end{equation}
and
\begin{equation}
\phi = \text{phase of } P_{10}\colon \ P_{10} = |P_{10}|e^{i\phi}.
\end{equation}
Equation \eqref{eq3.10} has time-varying coefficients and exact solutions of 
such equations are not always easy to come by. Here, we make another round of 
approximation, called the \emph{rotating wave approximation}, by first setting
\begin{equation}\label{eq3.17.1}
c_0(t) = C_0(t) e^{i\omega_0t},\quad c_1(t) = C_1(t)e^{i\omega_1t},
\end{equation}
and substituting them into \eqref{eq3.10} and then dropping terms involving 
$e^{\pm i(\omega+\nu)t}$, where $\omega\equiv \omega_0-\omega_1$ is called the 
atomic transition frequency. (From an experimental point of view, such $e^{\pm 
i(\omega+\nu)t}$ terms 
represent high frequency oscillations which cannot be observed in laboratory 
instrumentation. 
Furthermore, integrals involving highly oscillatory terms make 
insignificant contributions. In view of the fact that other non-resonant terms 
have already been discarded, these $e^{\pm i(\omega+\nu)t}$ terms representing 
highly non-resonant contributions should be neglected for consistency.) 
Then we obtain
\begin{equation}\label{eq3.12}
\left\{\begin{array}{l}
\dot c_0(t) = \dfrac{i\Omega_R}{2} e^{-i\phi} e^{i(\omega-\nu)t} c_1(t),\\
\noalign{\smallskip}
\dot c_1(t) = \dfrac{i\Omega_R}{2} e^{i\phi} e^{-i(\omega-\nu)t} c_0(t),
\end{array}\right.
\end{equation}
which in turn leads to a second order single ODE with constant coefficients
\begin{equation}\label{eq3.13}
\ddot c_0 - i(\omega-\nu) \dot c_0 + \frac{\Omega^2_R}{4} c_0 = 0.
\end{equation}
Solving for $c_0$ and $c_1$ in \eqref{eq3.12} and \eqref{eq3.13}, 
we obtain the 
explicit solution in terms of the initial condition $(c_0(0), c_1(0)$):
\begin{equation}
\left[\begin{matrix} c_0(t)\\ c_1(t)\end{matrix}\right] = \left[\begin{matrix}
e^{i\frac\Delta2 t} \left[\cos\left(\frac\Omega2 t\right) - i\frac\Delta\Omega 
\sin\left(\frac\Omega2 t\right)\right]&
e^{i\frac\Delta2 t} \cdot i \cdot \frac{\Omega_R}{\Omega} e^{-i\phi} \sin  
\left(\frac\Omega2 t\right)\\
e^{-i\frac\Delta2 t} \cdot i \frac{\Omega_R}\Omega e^{i\phi} \sin 
\left(\frac\Omega2 t\right)&
e^{-i\frac\Delta2 t} \left[\cos \left(\frac\Omega2 t\right) + 
i\frac\Delta\Omega 
\sin \left(\frac\Omega2 t\right)\right]\end{matrix}\right] \!
\left[\begin{matrix} c_0(0)\\ c_1(0)\end{matrix}\right],
\end{equation}
where $\Omega=\sqrt{\Omega_R^2+(\omega-\nu)^2}$.
At resonance,
\begin{equation}
\Delta = \omega-\nu =  0,
\end{equation}
and, thus
\begin{equation}
\frac\Delta\Omega = 0,\quad \Omega_R = \Omega.
\end{equation}
The above yields
\begin{equation}\label{eq3.13b}
\left[\begin{matrix} c_0(t)\\ c_1(t)\end{matrix}\right] = \left[\begin{matrix}
\cos\left(\frac\Omega2 t\right)&ie^{-i\phi} \sin\left(\frac\Omega2 t\right)\\
ie^{i\phi} \sin\left(\frac\Omega2 t\right)&\cos\left(\frac\Omega2 t\right)
\end{matrix}\right] \left[\begin{matrix} c_0(0)\\ c_1(0)\end{matrix}\right].
\end{equation}
(Note that if, instead, we relate $C_j(t)$ to $C_j(0)$ for $j=0,1$, then 
because of \eqref{eq3.17.1}, unlike the (special) unitary matrix in 
\eqref{eq3.13b} whose determine is 1, the corresponding unitary matrix will 
contain some additional phase factor(s).)
Write
\begin{align}\label{eq3.13a}
\theta &= \frac\Omega2 t\\
\phi' &= \phi+\pi,\nonumber
\end{align}
and re-name $\phi'$ as $\phi$. Then the above matrix becomes
\begin{equation}\label{eq3.14}
U_{\theta,\phi} \equiv \left[\begin{matrix}
\cos\theta&-ie^{i\phi} \sin\theta\\
-ie^{-i\phi}\sin\theta&\cos\theta\end{matrix}\right],
\end{equation}
which is the {\em 1-bit rotation unitary gate}.

Note that $\theta$ in \eqref{eq3.14} depends on $t$.
\medskip

\begin{rem}\label{rem3.1A}
If we write the Hamiltonian as (\cite[(5.2.44), p.~157]{SZ})
\begin{equation}\label{eq3.13c}
H = -\frac{\hbar \Omega}2 
[e^{-i\phi} |1\rangle\langle 0| + e^{i\phi} |0\rangle 
\langle 1|],
\end{equation}
then we obtain
\begin{equation}
e^{-\frac{i}\hbar Ht} = \left[\begin{matrix}
\cos\left(\frac\Omega2 t\right)&ie^{-i\phi} \sin \left(\frac\Omega2 t\right)\\
ie^{i\phi} \sin\left(\frac\Omega2 t\right)&\cos \left(\frac\Omega2 t\right)
\end{matrix}\right],
\end{equation}
which is the same as the matrix in \eqref{eq3.13b}. This Hamiltonian in 
\eqref{eq3.13c} is 
thus called the {\em effective (interaction) Hamiltonian\/} for the 2-level 
atom. It 
represents the essence of the original Hamiltonian in \eqref{eq3.3} after 
simplifying assumptions. From now on for the many physical systems under 
discussion, we will simply use the effective Hamiltonians for such systems as 
supported by theory and/or experiments, rather than to derive the Hamiltonians 
from scratch based on the  Schr\"odinger's equations coupled with the 
electromagnetic field. 
$\hfill\square$
\end{rem}

\subsection{Quantization of the Electromagnetic Field}\label{sec3.2}

The quantization of the electromagnetic radiation field as simple harmonic 
oscillators is 
important in quantum optics. This fundamental contribution is due to Dirac. 
Here 
we provide a motivation by following 
the approach of \cite[Chap.~1, pp.~3--4]{SZ}. 
We begin with the classical description of the field based on
Maxwell's equations.  These equations relate the electric and 
magnetic field
vectors $\pmb{E}$ and $\pmb{B}$, respectively.
Maxwell's equations lead to the following wave equation for the 
electric field:
\begin{equation}\label{3.2.4}
 \nabla^2 \pmb{E} - \frac1{c^2} \frac{\partial^2 \pmb{E}}{\partial t^2} 
   = 0, 
\end{equation}
along with a corresponding wave equation for the magnetic field.
The electric field has the spatial dependence
appropriate for a cavity resonator of length $L$.  We take the
electric field to be linearly polarized in the $x$-direction and 
expand in the
normal modes (so-called in the sense that they constitute orthogonal 
coordinates for oscillations) of the cavity
\begin{equation}\label{3.2.5}
 E_x(z,t)    = \sum_j A_j q_j (t) \sin (k_j z), 
\end{equation}
where $q_j$ is the normal mode amplitude with the dimension of a length,
$k_j = j \pi /L$, with $j = 1,2,3, \ldots$, and
\begin{equation}\label{3.2.6}
 A_j = \left(\frac{2\nu^2_j m_j}{V \epsilon_0}\right)^{1/2},
\end{equation}
with $\nu_j = j \pi c/L$ being the cavity eigenfrequency, $V = LA$ ($A$ is
the transverse area of the optical resonator) is the volume of the resonator
and $m_j$ is a constant with the dimension of mass.  The constant $m_j$ has
been included only to establish the analogy between the dynamical problem of a
single mode of the electromagnetic field and that of the simple harmonic
oscillator.  The equivalent mechanical oscillator will have a mass $m_j$, and
a Cartesian coordinate $q_j$.  The nonvanishing component of the magnetic
field $B_y$ in the cavity
is obtained from Eq.\ (\ref{3.2.5}):
\begin{equation}\label{3.2.7}
B_y = \sum_j A_j \left( \frac{\dot{q}_j \epsilon_0}{k_j} \right) \cos (k_j 
z).
\end{equation}

The classical Hamiltonian for the field is
\begin{equation}\label{3.2.8}
 H_{cl} = \frac{1}{2} \int_V d\tau(\epsilon_0 E^2_x + \frac{1}{\mu_0} 
B^2_y),
\end{equation}
where the integration is over the volume of the cavity.  It follows, on
substituting from  (\ref{3.2.5}) and (\ref{3.2.7}) for $E_x$ and $B_y$, 
respectively,
in  (\ref{3.2.8}), that
\begin{align}
 H_{cl} 
   &= \frac{1}{2} \sum_j (m_j \nu_j^2 q_j^2 + m_j \dot q_j^2) \nonumber \\
   &= \frac{1}{2} \sum_j \left(m_j \nu_j^2 q_j^2 + \frac{p_j^2}{ m_j} \right),
      \label{3.2.9}
\end{align}
where $p_j = m_j \dot q_j$ is the canonical momentum of the $j$th mode.
Equation (\ref{3.2.9}) expresses the Hamiltonian of the radiation field as a 
sum of independent oscillator energies.  
This suggests correctly that each mode of the field is 
dynamically equivalent to a mechanical harmonic oscillator.

The present dynamical problem can be quantized by identifying $q_j$ and
$p_j$ as operators which obey the commutation relations
\begin{align}%\label{3.2.10a}
 [q_j, p_{j^\prime}] 
   &= i \hbar \delta_{jj^\prime} ,\nonumber\\
\label{3.2.10b}
 [q_j, q_{j^\prime}] 
   &= [p_j, p_{j^\prime}] = 0.
\end{align}

In the following we shall restrict ourselves to a single mode of the radiation 
field modeled by a simple harmonic oscillator. 
The corresponding Hamiltonian is therefore given by
\begin{equation}\label{eq3.15}
H = \frac1{2m} p^2+ \frac12 m\nu^2 x^2,
\end{equation}
where
\begin{align}
\nonumber
p &= \text{the particle momentum operator } = \frac\hbar{i} \frac{d}{dx},\\
\nonumber
m &= \text{a constant with the dimensions of mass,}\\
\nonumber
x &= \text{the position operator (corresponding to the $q$ variable above),}\\
\nonumber
\nu &= \text{the natural (circular) frequency of the oscillator}\\
&\phantom{=}~\text{~~(a parameter related to the potential depth).}
\end{align}
So the eigenstates $\psi$ of the Schr\"odinger equation satisfy
\begin{equation}\label{eq3.16}
H\psi = -\frac{\hbar^2}{2m} \frac{d^2\psi}{dx^2} + \frac12 m\nu^2 x^2 \psi = 
E\psi.
\end{equation}

Let us  make a change of variables
\begin{equation}\label{eq3.16'}
y = \sqrt{\frac{m\nu}h}\ x,\quad \lambda = \frac{E}{\hbar \nu},
\end{equation}
then \eqref{eq3.16} becomes
\begin{equation}\label{eq3.17}
\frac12 \left[\frac{d^2\psi}{dy^2} - y^2\psi\right] = -\lambda\psi.
\end{equation}

We now define two operators
\begin{equation}
a = \frac1{\sqrt 2} (d/dy + y),\quad a^{\dag} = -\frac1{\sqrt 2} (d/dy-y).
\end{equation}
Note that $a^{\dag}$ is the Hermitian adjoint operator of $a$ with respect to 
the $L^2(\bb R)$ 
inner product. Then it is easy to check that for any sufficiently smooth 
function $\phi$ on $\bb R$,
\begin{equation}
(aa^{\dag}-a^{\dag}a)\phi = \phi,
\end{equation}
i.e.,
\begin{equation}\label{eq3.18}
\text{the commutator of $a$ with } a^{\dag} =[a,a^{\dag}] =\pmb{1},
\end{equation}
where $\pmb{1}$ denotes the identity operator but is often simply written as 
(the scalar) 1 in the 
physics literature. It is now possible to verify the following:

\n (i)\ Let $\widetilde H \equiv - \frac12\left(\frac{d^2}{dy^2}  - 
y^2\right)$. 
Then
\begin{equation}
\widetilde H  = a^{\dag}a + \frac12 \pmb{1} = aa^{\dag} - \frac12  \pmb{1}.
\end{equation}
(ii)\ If $\psi_\lambda$ is an eigenstate of $\widetilde H$ satisfying
\begin{equation}
\widetilde H \psi_\lambda = \lambda\psi_\lambda\qquad \text{(i.e., 
\eqref{eq3.17}),}
\end{equation}
then so are $a\psi_\lambda$ and $a^{\dag}\psi_\lambda$:
\begin{align*}
\widetilde H(a\psi_\lambda) &= \left(\lambda - 1\right) a\psi_\lambda,\\
\widetilde H(a^{\dag}\psi_\lambda) &= \left(\lambda + 1\right) a^{\dag}\psi_k.
\end{align*}
(iii)\ If $\lambda_0$ is the lowest eigenvalue (or energy level) 
of $\widetilde H$, then
\begin{equation}\label{eq3.18a}
\lambda_0 = \frac12.
\end{equation}
(iv)\ All of the eigenvalues of $\widetilde H$ are given by
\begin{equation}\label{eq3.18b}
\lambda_n = \left(n + \frac12\right),\qquad n=0,1,2,\ldots~.
\end{equation}
(v)\ Denote the $n$-th eigenstate $\psi_n$ by $|n\rangle$, for 
$n=0,1,2,\ldots$~. 
Then
\begin{equation}\label{eq3.19}
|n\rangle = \frac{(a^{\dag})^n}{(n!)^{1/2}} |0\rangle.
\end{equation}
Furthermore,
\begin{align}
\nonumber%\label{eq3.19a}
a^{\dag}a|n\rangle &= n|n\rangle,\qquad n=0,1,2,\ldots;\\
\nonumber%\label{eq3.20}
a^{\dag}|n\rangle &= \sqrt{n+1}\ |n+1\rangle,\qquad n=0,1,2,\ldots;\\
\label{eq3.21}
a|n\rangle &= \sqrt n\ |n-1\rangle,\qquad n=1,2,\ldots;\quad a|0\rangle = 0.
\end{align}
(vi) The completeness relation is
\begin{equation}
\pmb{1} = \sum^\infty_{n=0} |n\rangle\langle n|.
\end{equation}
(vii) The wave functions are
\begin{equation}
\psi_j(y) = N_jH_j(y) e^{-y^2/2},\qquad y 
= \left(\frac{m\nu}\hbar\right)^{1/2} 
x,
\end{equation}
$j=0,1,2,\ldots$, where $H_j(y)$ are the  Hermite polynomials of degree $j$, 
and 
$N_j$ is a normalization factor,

\begin{rem}\label{rem3.1}
\begin{itemize}
\item[(i)] The energy levels $\left(n+\frac12\right) \hbar\nu$ of $H$ (after 
converting back to the $x$-coordinate from the $y$-coordinate in 
\eqref{eq3.18b}) 
may be interpreted as the presence of $n$ quanta or photons of 
energy $\hbar \nu$. The eigenstates $|n\rangle$ are called the Fock states or 
the photon number states.
\item[(ii)] The energy level $E=\frac12 \hbar\nu$ (from \eqref{eq3.16'} and
\eqref{eq3.18a}) is called the ground state energy.
\item[(iii)] Because of \eqref{eq3.21}, we call $a$ and 
$a^{\dag}$, respectively, the annihilation and creation 
operators.$\hfill\square$
\end{itemize}
\end{rem}

\subsection{Cavity QED for the Quantum Phase Gate}\label{sec3.3}

\begin{figure}[ht]
\centerline{\epsfig{file=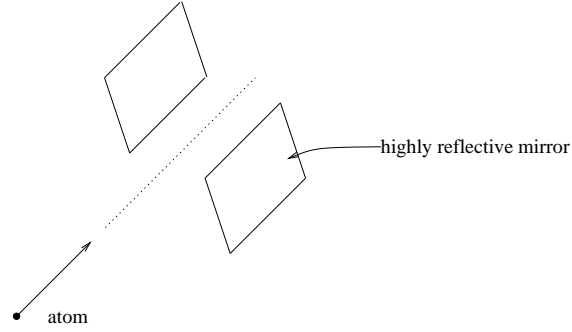,width=0.6\textwidth}}
\caption{\label{fig3.2}%
A Fabry--Perot cavity for electromagnetic radiation.
Inside the cavity, the mode structure of the radiation field is drastically
changed. In particular, there is a single privileged cavity mode that is 
resonant with the atomic transition and dominates the atom-field dynamics so
completely that the influence of all other modes is negligible. 
}  
\end{figure}

Cavity QED (quantum electrodynamics) is a system that enables the coupling 
of single atoms to only a few photons in a resonant cavity. It is realized by 
applying a large laser electric field in a narrow band of frequencies   
within a small Fabry--Perot cavity consisting of highly 
reflective mirrors. See a primitive drawing in Fig.~\ref{fig3.2}.

\begin{figure}
\centerline{\epsfig{file=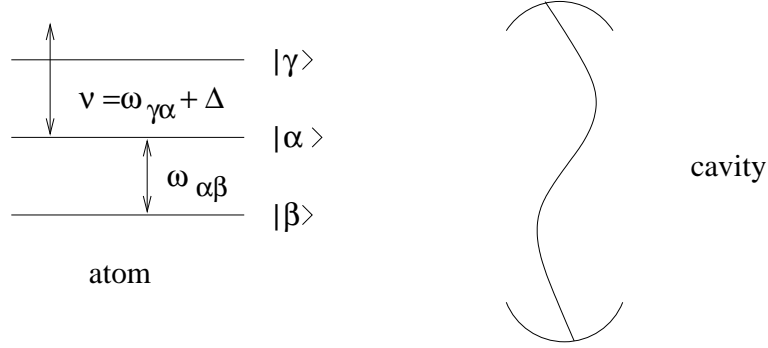,width=0.8\textwidth}}
\caption{\label{fig3.3}% 
Diagram for cavity QED.}
\end{figure}

A 3-level atom is injected into the cavity. An electron in the atom has three 
levels, $|\alpha\rangle$, $|\beta\rangle$ and $|\gamma\rangle$, as shown in 
Fig.~\ref{fig3.3}. 
Actually, the state $|\alpha\rangle$ will be used only as an {\em auxiliary\/} 
level because later we will define the states $|\beta\rangle$ and 
$|\gamma\rangle$ as the first qubit, $|1\rangle$ and $|0\rangle$, respectively. 
Once the atom enters 
the cavity, the strong electromagnetic field of the privileged cavity mode 
causes transitions of 
the electron between $|\alpha\rangle$ and $|\beta\rangle$, 
and a photon or photons are released or absorbed in this process. 
Of the photon 
states $|0\rangle$, $|1\rangle$, $|2\rangle,\ldots$ inside the cavity 
only $|0\rangle$ and $|1\rangle$ will be important 
(which physically mean 0 or 1 photo inside the cavity), and they define
the second qubit. 

The creation and  annihilation operators $a^{\dag}$ and $a$ act on the photon 
states 
$|n\rangle$ for $n=0,1,2,\ldots,\infty$, according to \eqref{eq3.21}.

The Hamiltonian for the atom-cavity field interaction is given by
\begin{equation}\label{eq3.22}
H = H_0 + H_1 + H_2,
\end{equation}
where
\begin{align}
\nonumber%\label{eq3.23}
H_0 &= \frac{\hbar \omega_{\alpha\beta}}{2} (|\alpha\rangle \langle\alpha| - 
|\beta\rangle\langle\beta|) = \text{ the atom's Hamiltonian,}\\
\label{eq3.24}
H_1 &= \hbar \nu a^{\dag}a 
= \text{ the Hamiltonian of the laser electric field of 
the cavity,}
\end{align}
and
\begin{align}
\label{eq3.25}
H_2 = \hbar g(|\alpha\rangle \langle \beta|a + |\beta\rangle\langle 
\alpha|a^{\dag}) 
=&\text{ the interaction Hamiltonian of}\\[-0.25\baselineskip]
&\text{ the laser field with the atom; with $g>0$.}\nonumber
\end{align}
See the derivation of \eqref{eq3.22} in \cite{Ra}. Note that the operators $a$ 
and $a^{\dag}$ appearing in \eqref{eq3.22} operate only on the second bit, 
while the 
rest of the operators operate only on the first bit. Also, operator 
$|\alpha\rangle \langle\alpha| - |\beta\rangle \langle\beta|$ 
can be written as 
$\sigma_z$; see equation \eqref{eq5.1a} later, because its matrix 
representation with 
respect to the ordered basis $\{|\alpha\rangle, |\beta\rangle\}$ is exactly 
$\sigma_z$.

\begin{lem}\label{lem3.1}
Let the underlying Hilbert space be
\begin{equation}\label{eq3.26}
\cl H = \text{span}\{|\alpha,n\rangle, |\beta,n\rangle, |\gamma,n\rangle| 
n=0,1,2,\ldots\}
\end{equation}
Then the Hamiltonian operator \eqref{eq3.22} has a family of 2-dimensional 
invariant subspaces
\begin{equation}\label{eq3.27}
V_n = \text{span}\{|\alpha,n-1\rangle, |\beta,n\rangle\},\qquad n=1,2,\ldots~.
\end{equation}
Indeed, with respect to the ordered basis in \eqref{eq3.27}, the matrix 
representation of $H$ on $V_n$ is given by
\begin{equation}\label{eq3.27a}
H|_{V_n} = \hbar \left[\begin{matrix}
\frac{1}{2}\omega_{\alpha\beta} + \nu(n-1)&g\sqrt n\\ 
g\sqrt n&-\frac{1}{2}\omega_{\alpha\beta} + \nu n\end{matrix}\right],\qquad 
n=1,2,3,\ldots~.
\end{equation}
\end{lem}

\begin{proof}
Even though the verification is straightforward, let us provide some details 
for 
the ease of future reference. We have, from \eqref{eq3.21}, and 
\eqref{eq3.22}--\eqref{eq3.25},
\begin{align}
H|\alpha,n-1\rangle &= H_0|\alpha,n-1\rangle + H_1|\alpha,n-1\rangle + 
H_2|\alpha,n-1\rangle\nonumber\\
&= \left[\frac{\hbar\omega_{\alpha\beta}}{2} |\alpha,n-1\rangle\right] 
+ [\hbar 
\nu(n-1)|\alpha,n-1\rangle] + [\hbar g\sqrt n|\beta,n\rangle]\nonumber\\
\label{eq3.27b}
&= \hbar \left[\frac{\omega_{\alpha\beta}}2 
+\nu(n-1)\right] |\alpha,n-1\rangle 
+ \hbar [g\sqrt n] |\beta,n\rangle.
\end{align}
Similarly,
\begin{equation}\label{eq3.27c}
H|\beta,n\rangle= \hbar \left[-\frac{\omega_{\alpha\beta}}2 + \nu n\right] 
|\beta,n\rangle + \hbar [g\sqrt n] |\alpha, n-1\rangle.
\end{equation}
Therefore, we obtain \eqref{eq3.27a}.
\end{proof}

\begin{lem}\label{lem3.2}
On the invariant subspace $V_n$ in \eqref{eq3.27}, 
the Hamiltonian operator $H$ 
has two eigenstates
\begin{equation}\label{eq3.28}
\left\{\begin{array}{l}
|+\rangle_n \equiv \cos \theta_n|\alpha,n-1\rangle - 
\sin\theta_n|\beta,n\rangle,\\[1ex]
|-\rangle_n \equiv \sin\theta_n|\alpha,n-1\rangle + 
\cos\theta_n|\beta,n\rangle,
\end{array}\right.
\end{equation}
where 
\begin{align}
\sin\theta_n &\equiv \frac{\Omega_n-\Delta}D,\quad \cos \theta_n \equiv 
\frac{2g\sqrt n}D,\nonumber\\
D &\equiv [(\Omega_n-\Delta)^2 + 4g^2n]^{1/2},\nonumber\\
\Omega_n &\equiv (\Delta^2 +4g^2n)^{1/2},\nonumber\\
\label{eq3.28a}
\Delta &\equiv \nu -\omega_{\alpha\beta} = \text{ the detuning frequency,}
\end{align}
with eigenvalues (i.e., energy levels)
\begin{equation}\label{eq3.29}
E_{\pm(n)} = \hbar\left[n\nu + \frac12(-\nu \mp \Omega_n)\right],
\end{equation}
such that
\begin{equation}\label{eq3.30}
H|+\rangle_n = E_{+(n)} |+\rangle_n,\quad H|-\rangle_n = E_{-(n)}|-\rangle_n.
\end{equation}
\end{lem}

\begin{proof}
These eigvenvalues and eigenvectors can be computed in a straightforward way, 
using the $2\times 2$ matrix representation of $H$ on $V_n$, \eqref{eq3.27a}, 
with respect to the ordered 
basis of $V_n$ chosen as in \eqref{eq3.27}.
\end{proof}

\begin{rem}\label{rem3.1a}
The states $|+\rangle_n$ and $|-\rangle_n$ in \eqref{eq3.28} are called the 
{\em dressed states\/} in the sense that atoms are dressed by electromagnetic 
fields. 
These two states are related to the splitting of the spectral 
lines due to the electric field. 
$\hfill\square$
\end{rem}
\medskip

Now, let us assume {\em large detuning}:
\begin{equation}\label{eq3.31}
|\Delta| \gg 2g\sqrt n~.
\end{equation}
Also, $|\Delta|$ remains small in comparison to the 
transition frequency so that approximate resonance is maintained. 
Accordingly, this is satisfied only for small values of $n$.
Then
\begin{align}\nonumber
\Omega_n &= (\Delta^2+4g^2n)^{1/2} = \Delta\left(1 + 
\frac{2g^2n}\Delta\right)^{1/2} \approx \Delta + \frac{2g^2n}\Delta;\\
\sin\theta_n &= (\Omega_n-\Delta)/D\approx 0;\quad \cos\theta_n = 2g\sqrt 
n/D\approx 1.
\end{align}
Thus, from \eqref{eq3.28},
\begin{equation}\label{eq3.32}
|+\rangle_n \approx |\alpha,n-1\rangle, \quad |-\rangle_n\approx 
|\beta,n\rangle.
\end{equation}
\medskip

\begin{rem}\label{rem3.2}
By the assumption \eqref{eq3.30} of large detuning, from \eqref{eq3.29} using 
\eqref{eq3.28a} and \eqref{eq3.31}, we now have
\begin{align}
E_{+(n)} &= \hbar\left[n\nu + \frac12 (-\nu-\Omega_n)\right]\nonumber\\
&\approx \hbar\left[\left(n-\frac12\right)\nu - \frac12 \left(\Delta + 
\frac{2g^2n}\Delta\right)\right]\nonumber\\
&= \hbar\left[\left(n-\frac12\right)\nu - \frac12 (\nu-\omega_{\alpha\beta}) - 
\frac{g^2n}\Delta\right]\nonumber\\
\label{eq3.32a}
&= \hbar \left[\frac{\omega_{\alpha\beta}}2 + \nu(n-1)\right] - \frac{\hbar 
g^2n}\Delta.
\end{align}
Similarly, we have
\begin{equation}\label{eq3.32b}
E_{-(n)} \approx \hbar \left[-\frac{\omega_{\alpha\beta}}2 +\nu n\right] + 
\frac{\hbar g^2n}\Delta,
\end{equation}
such that
\begin{align}\nonumber
H|+\rangle_n &\approx H|\alpha,n-1\rangle = E_{+(n)}|\alpha,n-1\rangle,\\
H|-\rangle_n &\approx H|\beta,n\rangle = E_{-(n)} |\beta,n\rangle,
\end{align}
where $E_{\pm(n)}$ are now given by \eqref{eq3.32a} and \eqref{eq3.32b}.

Thus, we see that the assumption \eqref{eq3.30} of large detuning causes the 
off-diagonal terms in the matrix \eqref{eq3.27a} (which are contributed by the 
{\em interaction Hamiltonian\/} $H_2$ in \eqref{eq3.24}) to disappear. The 
de facto or, {\em effective interaction Hamiltonian}, has now become 
\begin{equation}\label{eq3.32c}
\widetilde H_2 
= -\frac{\hbar g^2}\Delta (aa^{\dag}|\alpha\rangle\langle\alpha| 
-a^{\dag}a|\beta\rangle\langle\beta|),
\end{equation}
such that $\widetilde H \equiv H_0 + H_1 + \widetilde H_2$ now admits a 
diagonal 
matrix representation
\begin{equation}\label{eq3.32d}
\widetilde H|_{V_n} = \left[\begin{matrix} E_{+(n)}&0\\ 0&E_{-(n)}\end{matrix} 
\right]
\end{equation}
with respect to the ordered basis $\{|\alpha,n-1\rangle, |\beta,n\rangle\}$ of 
$V_n$, with $E_{\pm(n)}$ here given by \eqref{eq3.32a} and \eqref{eq3.32b}. 
$\hfill\square$
\end{rem}
\medskip

We now define the first qubit by
\begin{equation}\label{eq3.33}
|\beta\rangle = |1\rangle,\quad |\gamma\rangle = |0\rangle,
\end{equation}
where $|\gamma\rangle$ is the state in Fig.~\ref{fig3.3} 
that is {\em detached\/} (or 
{\em off-resonance\/}) from the Hamiltonian $H$, i.e.,
\begin{equation}\label{eq3.34}
H|\gamma\rangle = 0.
\end{equation}

For the second qubit, we choose $n=1$ for $V_n$ in \eqref{eq3.27} and then see 
that the 
second bit is only 0 or 1, i.e., the cavity can have 0 or 1 photon. The cavity 
field is nearly resonant with the $|\alpha\rangle \leftrightarrow 
|\beta\rangle$ 
transition: $\nu = \omega_{\alpha\beta} + \Delta$.

\begin{thm}\label{thm3.3}
Let the effective total Hamiltonian be
\begin{align}
\widetilde H = H_0 + H_1 + \widetilde H_2= \hbar& 
\left[\frac{\omega_{\alpha\beta}}2 (|\alpha\rangle \langle\alpha| - 
|\beta\rangle \langle\beta|) + \nu a^{\dag}a\right.\nonumber\\
\label{eq3.35}
&\enskip \left. - \frac{\hbar g^2}\Delta (aa^{\dag} 
|\alpha\rangle\langle\alpha| - a^{\dag}a|\beta\rangle \langle\beta|\right].
\end{align}
Then
\begin{equation}\label{eq3.36}
\widetilde V \equiv \text{span}\{|0,0\rangle, |0,1\rangle, |1,0\rangle, 
|1,1\rangle\}
\end{equation}
is an invariant subspace of $\widetilde H$ in the underlying Hilbert space 
\eqref{eq3.26}, such that
\begin{equation}\label{eq3.37}
\left\{\begin{array}{l}
\widetilde H|0,0\rangle = 0,\quad \widetilde H |0,1\rangle = 0,\quad 
\widetilde 
H|1,0\rangle = 0\\  \displaystyle
\widetilde H|1,1\rangle = E_{-(1)} |1,1\rangle, \text{ where } E_{-(1)} 
= \hbar 
\left(-\frac{\omega_{\alpha\beta}}2 + \nu + \frac{\hbar g^2}\Delta\right).
\end{array}\right.
\end{equation}
Consequently, with respect to the ordered basis of $\widetilde V$ in 
\eqref{eq3.36}, $\widetilde H$ admits a diagonal matrix representation
\begin{equation}\label{eq3.38}
\widetilde H|_{\widetilde V} = \left[\begin{matrix}
0&&&&\bigcirc\\
&0\\
&&&0\\
&\bigcirc&&&E_{-(1)}\end{matrix}\right],
\end{equation}
with the evolution operator
\begin{equation}\label{eq3.39}
e^{-i\widetilde Ht/\hbar}|_{\widetilde V} = \left[\begin{matrix}
1\\ &&&&&\bigcirc\\ &&1\\ &&&&1\\ &\bigcirc\\
&&&&&\exp\left(-i\frac{E_{-(1)}t}\hbar \right)\end{matrix}\right].
\end{equation}
\end{thm}

\begin{proof}
We have
\begin{equation}
\widetilde  H|0,0\rangle = \widetilde H|\gamma,0\rangle = 0\quad 
\text{and}\quad 
\widetilde H|0,1\rangle = \widetilde H|\gamma,1\rangle = 0, \text{ by 
\eqref{eq3.34}.}
\end{equation}
Also,
\begin{equation}
\widetilde H|1,0\rangle = \widetilde H|\beta,0\rangle = 0, \text{ by 
\eqref{eq3.32d},}
\end{equation}
where $V_n$ is chosen to be $V_0$ by setting $n=0$ therein. But 
\begin{equation}
\widetilde H|1,1\rangle = \widetilde H|\beta,1\rangle = E_{-(1)} |1,1\rangle
\end{equation}
by \eqref{eq3.32d} where $V_n$ is chosen to be $V_1$ by setting $n=1$ therein. 
So both \eqref{eq3.38} and \eqref{eq3.39} follow.
\end{proof}

The unitary operator in \eqref{eq3.39} gives us the quantum phase gate
\begin{equation}\label{eq3.40}
Q_\eta = \left[\begin{matrix}
1\\ &1&&&\bigcirc\\  &&&1\\ &\bigcirc&&&e^{i\phi}\end{matrix}\right]
\end{equation}
with
\begin{equation}\label{eq3.41}
\phi \equiv - \frac{E_{-(1)}}\hbar t.
\end{equation}

We can now invoke Theorem \ref{thm2.2} in Section \ref{sec2} to conclude the 
following.

\begin{thm}\label{thm3.4}
The collection of 1-bit gates $U_{\theta,\phi}$ in \eqref{eq3.14} from the 
2-level atoms and the 2-bit gates $Q_\eta$ in \eqref{eq3.40} from cavity QED 
is universal for quantum computation.$\hfill\square$
\end{thm}

\section{Ion Traps}\label{sec4}

Ion traps utilize charged atoms with internal electronic states to represent 
qubits. Ion traps isolate and confine small numbers of charged atoms by means 
of electromagnetic fields. The ions are then cooled using laser beams until 
their kinetic energy is much smaller than the inter-ionic potential energy. 
Under these conditions, the ions form a regular array in the trap. Laser beams 
can be tuned to excite the electronic states of particular ions and to couple 
these internal states to the center-of-mass (CM) vibrational motion of the ion 
array. This coupling provides entanglement for quantum computation.

\begin{figure}[t]
\centerline{\epsfig{file=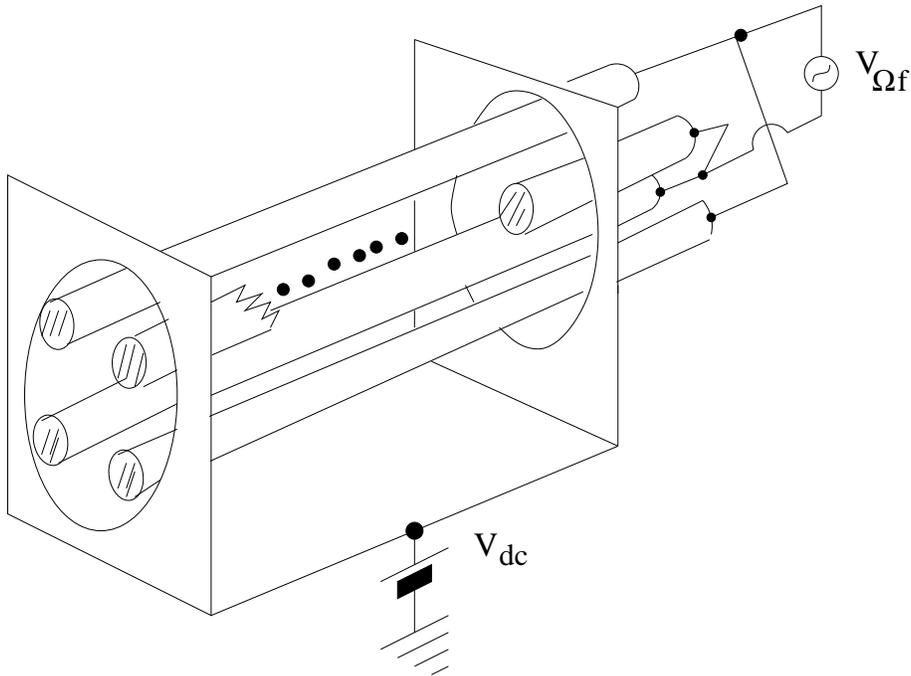,width=0.95\textwidth}}
\caption{\label{fig4.1}%
Diagram of a linear ion trap, showing the radio-frequency 
voltage connections on the four rods of the trap, and the dc voltage 
connection 
for an axial dc potential well. One rod is partially cut back to reveal 6 cold 
ions (hiding behind, not inside, the rod) confined in a linear array within
the trap. 
The ions would be visible due to scattered light, if they were all 
simultaneously excited by a laser beam tuned to an internal resonance 
transition, at a wavelength in the visible region of the spectrum. The 
separation of the trap rods is on the order of a millimeter.
}
\end{figure}

A Paul, or radio-frequency, ion trap confines ions by the time average of an 
oscillating force $F(\pmb{r},t)$, where $\pmb{r} = (x,y,z)$, arising from a 
non-uniform electric field $E(\pmb{r}) \cos(\mu t)$ acting on the ion charge 
$e$ with mass $m$. 
In one dimension, $\langle F(X,t)\rangle = -\partial\langle 
U(X,t)\rangle/\partial X$, where $\langle U(X,t)\rangle$ is an effective 
time-averaged potential energy, and  
$U(X,t) = e^2E^2(X)\cos^2(\Omega t)/(m\Omega^2)$. The time 
average 
is denoted by $\langle~~\rangle$ and here, it is defined by $\langle  
y(t)\rangle = \frac\Omega{2\pi} \int^{\frac{2\pi}\Omega}_0 y(t)dt$, where 
$y(t)$ 
is 
periodic with period $2\pi/\Omega$. $X$ describes the (relatively slow) motion 
of a 
guiding center, about which the ions execute small oscillations at frequency 
$\Omega$. The trap is constructed so that the guiding center motion in two 
dimensions (say $x$ and $y$) is harmonic, with frequencies $\nu_x$ and $\nu_y$ 
which are much smaller than the frequency $\Omega$. The trap electrodes may 
confine 
the ions in a circle or other closed 2-dimensional geometry, or a linear trap 
may be produced by using a dc potential to confine the ions in the third 
spatial dimension $z$; see Fig.~\ref{fig4.1}.

\begin{figure}[ht]
\centerline{\epsfig{file=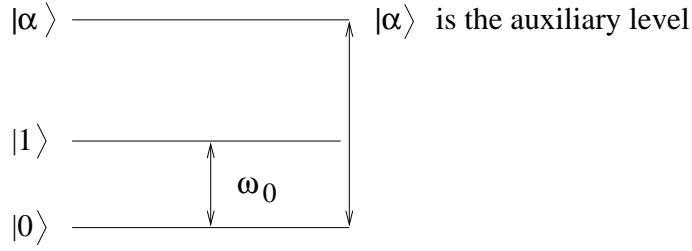,height=1.3in}}
\caption{\label{fig4.2}%
Diagram of the energy levels of an idealized ion, which might 
serve as the two qubit levels, plus an auxiliary level used in a particular 
implementation of a quantum circuit. The actual levels of a real ion are much 
more complex, but approximate the ideal case when the exciting lasers are 
properly tuned.}
\end{figure}

Once confined in the trap, the ions may be cooled using laser light, so that 
they condense into a linear array along the trap axis. Each ion oscillates 
with small amplitude about the zero of the time-averaged radial potential. 
The electrostatic repulsion of the ions along the linear axis of the trap 
combats a spatially varying dc potential applied for axial confinement, 
so that below milliKelvin $(1\times 10^{-3} K)$ temperatures, 
the ions in the array are 
separated by several micrometers $(10^{-6}m)$. Under these conditions, the ion 
confinement time is very long, many hours or days, and each ion can be 
separately excited by a focused laser beam, so that long-lived internal levels 
of each ion can serve as the 
two states of a qubit (Fig.~\ref{fig4.2}). 
Regarding the axial kinetic motion, the ions 
oscillate axially as a group, so collective oscillations such as the CM 
oscillations couple the individual ions, or qubits, permitting the 
implementation of multi-qubit gates based on quantum entanglement of the CM 
motion with each laser-excited qubit state. There are many CM modes of 
oscillation, depending on the oscillations of individual ions or groups of 
ions, 
each with a characteristic frequency $\omega_{\text{CM}}$.

\begin{figure}[t]
\begin{picture}(300,240)(30,0)
\put(97,150){\epsfig{file=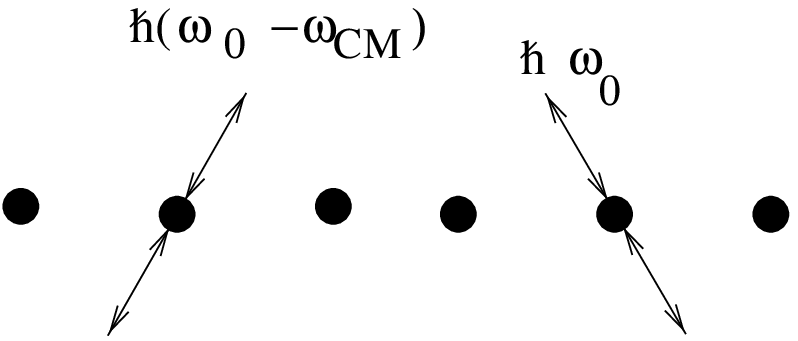,height=1in}}
\put(60,0){\epsfig{file=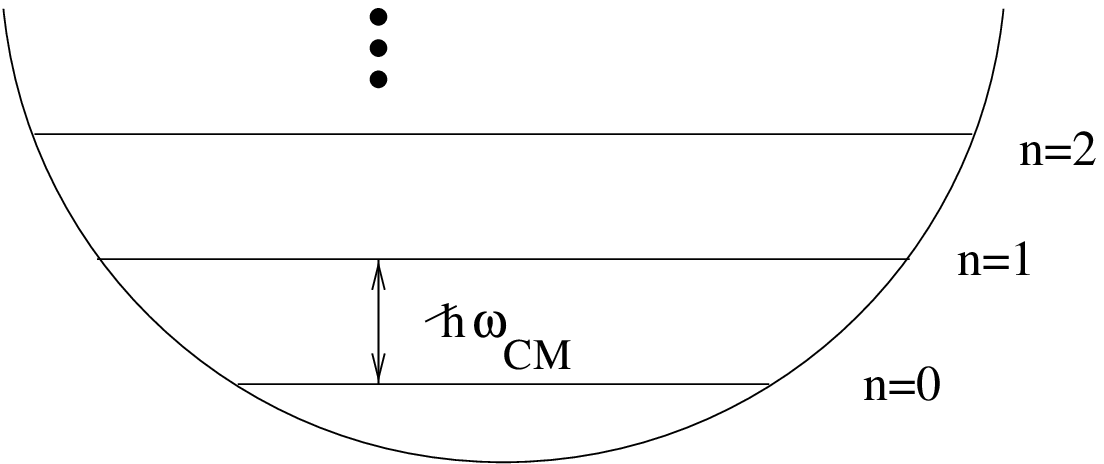,height=1.5in}}
\put(35,230){\textsf{(a)}}
\put(35,130){\textsf{(b)}}
\end{picture}
\caption{\label{fig4.3}% 
(a)~A diagram showing 6 cold ions in a linear array. the 
arrows show two of the ions individually excited by focused laser beam 
standing waves. One laser is tuned to the internal qubit transition frequency 
$\omega_0$, while the other laser is tuned to the first red sideband frequency 
$\omega_0-\omega_{\text{CM}}$, which couples the internal transition to the 
center-of-mass vibration of the ion array at $\omega_{\text{CM}}$ in the ion 
trap.\newline
(b)~A diagram of the harmonic potential well of the axial ion 
center-of-mass vibration, showing the lowest energy levels. 
The six ions in (a) would populate one of these energy levels, 
acting as a single entity with the CM coordinates.
}
\end{figure}

A prototype scheme was proposed by Cirac and Zoller \cite{CZ}, which is based 
on a confined ion system as outlined above. Assume the quantized axial CM
normal mode oscillation has frequency $\omega_{\text{CM}}$ and discrete
levels $n\hbar \omega_{\text{CM}}$ with quantum number $n=0,1,2,\ldots$ (the
ground state zero-point energy $\hbar \omega_{\text{CM}}/2$ in Remark
\ref{rem3.1}(ii) is disregarded);  see Fig.~\ref{fig4.3}(b). 
A quantum of 
oscillation $\hbar\omega_{\text{CM}}$ is called a {\em phonon}. Each ion is 
individually excited by focused standing wave laser beams. Two internal 
hyperfine structure states of the ion (i.e., states coupling the electrons and 
the nucleus of the ion) in the ground term (or lowest electronic set of 
states), 
separated by the frequency $\omega_0$, can be identified as the qubit states 
$|0\rangle$ and $|1\rangle$. Alternatively, a ground level and  a metastable 
excited level might be used. 
By thus coupling only these selected states of the ion by electromagnetic
fields we are  
led to a {\em 2-level atom\/} analogously as in Section \ref{sec3}. These 
states can be coupled using electric dipole transitions to higher-lying
states,  
and there is also an internal auxiliary level, denoted $|\alpha\rangle$, which 
can be coupled to the qubit states; see Fig.~\ref{fig4.2}. 
Consequently a linear $N$-ion 
system can be specified in terms of quantum numbers by the wavefunction 
$|\psi\rangle_N = |j_1\rangle |j_2\rangle\cdots |j_N\rangle 
|k\rangle_{\text{CM}}$, 
where states (of each ion) $j_1,j_2,\ldots, j_N\in \{0,1\}$ and 
$k=0,1,2,\ldots, 
N-1$ describes the populated quantum state of the CM motion. The system is 
initialized with $k=0$ when the ions are very cold, and with all qubit states 
$j_1,j_2,\ldots, j_N$ set to 0 by  inducing appropriate laser transitions in 
the individual ions.

Fig.~4.3(b) gives a diagram of the harmonic potential well of the axial ion 
center-of-mass vibration, showing the lowest energy levels. The six ions in 
Fig.~\ref{fig4.3}(a) would populate one of these energy levels, 
acting as a single entity with the CM coordinates.

The acting laser beam tuned to frequency $\omega_0$ causes a dipole 
transition coupling the two qubit levels $|0\rangle$ and $|1\rangle$ in 
Fig.~\ref{fig4.2}. 
Using the same rotating wave approximation as in Section \ref{sec3}, we obtain 
the 1-bit rotation matrix $U_{\theta,\phi}$ as in \eqref{eq3.14}.

Therefore, according to the theory of universal quantum computing (Theorem 
\ref{thm2.1}) 
in Section \ref{sec2}, now all we need to show is an entanglement operation 
between two qubits.

We assume that two ions corresponding to the two qubits are coupled 
in the {\em Lamb--Dicke regime}, i.e., 
the amplitude of the ion oscillation in the ion trap 
potential is small compared to the wavelength of the incident laser beam. This 
is expressed in terms of the Lamb--Dicke criterion:
\begin{equation}\label{eq4.1}
\eta(\Omega/2\omega_{\text{CM}}) \ll 1
\end{equation}
where

\begin{equation}
  \begin{array}{rl }
\eta \equiv k z_0\,:& \text{the Lamb--Dicke parameter,}\\
z_0\,:& \text{a ``zero-point'' oscillation amplitude,}\\
& \text{at the lowest energy of CM vibration,}\\
k = 2\pi/\lambda \,:& \text{the wave number of the laser radiation,}\\
\omega_{\text{CM}} \,:&  \text{the frequency of the CM motion referred to 
earlier,}\\
\Omega\,:&\text{the Rabi frequency (cf.\ \eqref{eq3.11}).}
\end{array}
\end{equation}

\begin{figure}[t]
\centerline{\epsfig{file=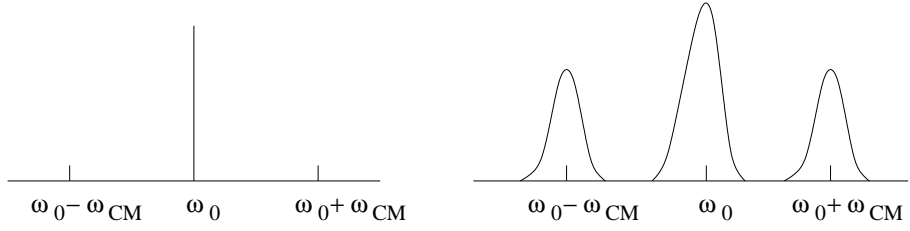,width=0.95\textwidth}}
\caption{\label{fig4.4}% 
$\omega_0-\omega_{\text{CM}}$ is the red sideband frequency 
which 
may be excited to entangle the qubit with the CM motion. The right figure 
indicates that when the 
Lamb--Dicke criterion is satisfied the sideband frequencies $\omega_0 \pm 
\omega_{\text{CM}}$ and $\omega_0$, which have finite widths due to the 
Heisenberg 
uncertainty principle, are resolved, enabling separate excitations by the 
laser beam. (The frequency $\omega_0+\omega_{\text{CM}}$ is called the blue 
sideband frequency; see \cite{WMILKM}.)}
\end{figure}

When \eqref{eq4.1} is satisfied, the ``sideband'' frequencies $\omega_0\pm 
\omega_{\text{CM}}$ arising from the CM oscillation in the laser standing wave 
are {\em resolved\/} (see Fig.~4.4 and the captions therein), 
and can be 
separately excited by the laser beam; see Fig.~\ref{fig4.3}(a). 
Then the interaction
Hamiltonian coupling the internal qubit states of each ion $j$, $j=1,\ldots, 
N$, to the CM motion is similar to \eqref{eq3.25}, with coupling constant $g = 
\frac{\eta\Omega}{2\sqrt N}$:
\begin{equation}\label{eq4.2}
H_j = \frac{\hbar\eta\Omega}{2\sqrt N} 
[|1\rangle_j \langle 0|_j e^{-i\phi} a + 
|0\rangle_j \langle 1|_j e^{i\phi} a^{\dag}],\qquad j=1,\ldots, N,
\end{equation}
where $a$ and $a^{\dag}$ are, respectively, the annihilation and creation 
operators 
(see Subsection \ref{sec3.2}) for the phonons defined by
\begin{align}\nonumber
a|n\rangle_{\text{CM}} &= \sqrt n\ |n-1\rangle_{\text{CM}}, \text{ for } 
n=1,2,\ldots;\quad a|0\rangle_{\text{CM}} = 0|0\rangle_{\text{CM}} = 0,\\
a^{\dag}|n\rangle_{\text{CM}} &= \sqrt{n+1}\ |n+1\rangle_{\text{CM}}, \text{ 
for } 
n=0,1,2,\ldots; \text{ cf.\  \eqref{eq3.21},}
\end{align}
and
\begin{align}\nonumber
\phi = &\text{ the angle in the spherical coordinate system $(r,\theta,\phi)$ 
with respect}\\[-0.25\baselineskip]
&\text{ to the $x$-axis of the laboratory coordinate frame.}
\end{align}
The derivations of \eqref{eq4.2} may be found in \cite[Section 5.1]{Hu}
 where the 
authors give an argument based on normal modes of phonons. At low enough 
temperature (i.e., the Lamb--Dicke regime), only the lowest order or the CM mode 
plays a role, and this leads to the Hamiltonian \eqref{eq4.2} used by Cirac and 
Zoller in \cite{CZ}.

\begin{rem}\label{rem4.1}
It can be shown that the Rabi frequency $\Omega_R$ for the internal qubit 
transition coupled by $a$ and $a^{\dag}$ to the vibration is 
$\eta\Omega/(2\sqrt 
N)$, 
much smaller than that, $\Omega$, in (\ref{eq3.11}) for a single qubit 
(rotation) 
operation.$\hfill\square$
\end{rem}

\begin{lem}\label{lem4.1}
Let $H_j$ be given by \eqref{eq4.2} for $j=1,2,\ldots, N$. Then $H_j$ has a 
family of invariant two-dimensional subspaces
\begin{equation}
V_{j,k} \equiv \text{ span}\{|0\rangle_j |k+1\rangle_{\text{CM}}, |1\rangle_j 
|k\rangle_{\text{CM}}\}, \quad k=0,1,2,\ldots,
\end{equation}
in the Hilbert space
\begin{equation}
V_j \equiv \text{ span}\{|0\rangle_j |k\rangle_{\text{CM}}, 
|1\rangle_j|k\rangle_{\text{CM}}\mid k=0,1,2,\dots\}.
\end{equation}
On the orthogonal complement of all the $V_{j,k}$'s, i.e., $\widetilde V_j = 
\left(\text{span } \bigcup\limits^\infty_{k=0} V_{j,k}\right)^\bot \subset 
V_j$, 
the action of $H_j$ is $0\colon \ H_j |\psi\rangle = 0$ for all $\psi\in 
\widetilde V_j$.
\end{lem}

\begin{proof}
We have
\begin{align}\nonumber
H_j|0\rangle_j|k+1\rangle_{\text{CM}} &= \frac{\hbar\eta\Omega}{2\sqrt N} 
[|1\rangle_j \langle 0|0\rangle_j e^{-i\phi} (a|k+1\rangle_{\text{CM}}) + 
|0\rangle_j \langle 1|0\rangle_j e^{i\phi}(a^{\dag}|k+1\rangle_{\text{CM}})]\\
&= \frac{\hbar\eta\Omega}{2\sqrt N} e^{-i\phi} \sqrt{k+1}\ |1\rangle_j 
|k\rangle_{\text{CM}},
\end{align}
and, similarly,
\begin{equation}
H_j|1\rangle_j |k\rangle_{\text{CM}} = \frac{\hbar\eta\Omega}{2\sqrt N} 
e^{i\phi} \sqrt{k+1}\ |0\rangle_j |k+1\rangle_{\text{CM}}.
\end{equation}
Thus, $V_{j,k}$ is an invariant two-dimensional subspace of $H_j$ in $V_j$.
\end{proof}

Let the time-evolution operator of $H_j$ (depending on $\phi$) on $V_j$ be 
$U_j(t,\phi) = e^{-\frac{i}{\hbar} H_jt}$. Then on the subspace $V_{j,k}$, 
$k=0,1,2,\ldots,\infty$, with respect to the ordered basis $\{|0\rangle_j 
|k+1\rangle_{\text{CM}}, |1\rangle_j |k\rangle_{\text{CM}}\}$, $H_j$ 
admits the following matrix representation
\begin{equation}
H_j = \frac{\hbar\eta\Omega}{2} \sqrt{\frac{k+1}N} \left[\begin{matrix}
0&e^{-i\phi}\\ e^{i\phi}&0\end{matrix}\right].
\end{equation}
Thus, with $H_j$ restricted to $V_{j,k}$, its time-evolution operator is given 
by
\begin{align}
U_{j,k}(t,\phi) &\equiv U_j(t,\phi)\big|_{V_{j,k}} = e^{-\frac{i}{\hbar} 
H_jt}\Big|_{V_{j,k}}\nonumber\\
\label{eq4.3}
&= \left[\begin{matrix} \cos(\cl E_kt)&-ie^{-i\phi} \sin (\cl E_kt)\\
-ie^{i\phi}\sin(\cl E_kt)&\cos(\cl E_kt)\end{matrix}\right],\quad \cl E_k 
\equiv 
\frac{\hbar\eta\Omega}{2} \sqrt{\frac{k+1}N}.
\end{align}

Also, note that
\begin{equation}\label{eq4.4}
U_j(t,\phi)|\psi\rangle = |\psi\rangle\quad \text{for all}\quad |\psi\rangle 
\in 
\widetilde V_j,
\end{equation}
because the action of $H_j$ on $\widetilde V_j$ is annihilation. In physics, 
these states in $\widetilde V_j$ are said to be {\em off-resonance}.

Next,  define
\begin{equation}\label{eq4.5}
H^{\mathrm{aux}}_j 
= \frac{\hbar\eta\Omega}{2\sqrt N} [|\alpha\rangle_j \langle 
0|_j e^{-i\phi}a + |0\rangle_j \langle\alpha|_j e^{i\phi}a^{\dag}],\qquad 
j=1,2,
\end{equation}
where $|\alpha\rangle$ is the auxiliary quantum state 
as indicated in Fig.~\ref{fig4.2}. 
Then, similarly to Lemma \ref{lem4.1}, we have the following.

\begin{lem}\label{lem4.2}
Let $H^{\mathrm{aux}}_j$ be given by \eqref{eq4.5} for $j=1,2,\ldots, N$. Then 
$H^{\mathrm{aux}}_j$ has a family of invariant two-dimensional subspaces
\begin{equation}
V^{\mathrm{aux}}_{j,k} \equiv 
\text{span}\{|0\rangle_j |k+1\rangle_{\text{CM}}, 
|\alpha\rangle_j |k\rangle_{\text{CM}}\},\qquad k=0,1,2,\ldots,
\end{equation}
in the Hilbert space
\begin{equation}
V^{\mathrm{aux}}_j \equiv \text{span}\{|0\rangle_j |k\rangle_{\text{CM}}, 
|\alpha\rangle_j |k\rangle_{\text{CM}}\mid k=0,1,2,\ldots\}.
\end{equation}
On the orthogonal complement $\widetilde V^{\mathrm{aux}}_j \equiv 
\left(\text{span } 
\bigcup\limits^\infty_{k=0} V^{\mathrm{aux}}_{j,k}\right)^\bot 
\subset V^{\mathrm{aux}}_j$, the action of $H^{\mathrm{aux}}_j$ is $0\colon \ 
H^{\mathrm{aux}}_j |\psi\rangle = 0$ for all $|\psi\rangle \in \widetilde 
V^{\mathrm{aux}}_j$.$\hfill\square$
\end{lem}

The time-evolution operator corresponding to $H^{\mathrm{aux}}_j$ on 
$V^{\mathrm{aux}}_j$ is denoted as 
$U^{\mathrm{aux}}_j(t,\phi) = e^{-\frac{i}\hbar 
H^{\mathrm{aux}}_jt}$. Now, restrict $H^{\mathrm{aux}}_j$ to the invariant 
two-dimensional subspace $V^{\mathrm{aux}}_{j,k}$ with ordered basis 
$\{|0\rangle_j |k+1\rangle_{\text{CM}}, |\alpha\rangle_j 
|k\rangle_{\text{CM}}\}$; its evolution operator has the matrix representation
\begin{align}\nonumber
U^{\mathrm{aux}}_{j,k}(t,\phi) &\equiv 
U^{\mathrm{aux}}_j(t,\phi)\big|_{V^{\mathrm{aux}}_{j,k}} = e^{-\frac{i}\hbar 
H^{\mathrm{aux}}_jt}\big|_{V^{\mathrm{aux}}_{j,k}}\\
&= \left[\begin{matrix} \cos(\cl E_kt)&-ie^{-i\phi} \sin (\cl E_kt)\\
-ie^{i\phi}\sin(\cl E_kt)&\cos(\cl E_kt)\end{matrix}\right],\quad \cl E_k 
\equiv 
\frac{\hbar\eta\Omega}{2} \sqrt{\frac{k+1}N}.\label{eq4.6}
\end{align}
Here, we again also have
\begin{equation}\label{eq4.7}
U^{\mathrm{aux}}_j (t,\phi)|\psi\rangle 
= |\psi\rangle \quad \text{for all}\quad 
|\psi\rangle\in \widetilde V^{\mathrm{aux}}_j,
\end{equation}
because $H^{\mathrm{aux}}_j$ annihilates the subspace $\widetilde 
V^{\mathrm{aux}}_j$.
Using the CM mode as the bus, we can now derive the following 2-bit quantum 
phase gate.

\begin{thm}\label{thm4.3}
Let $U_j(t,\phi), U_{j,k}(t,\phi), U^{\mathrm{aux}}_j(t,\phi)$ and 
$U^{\mathrm{aux}}_{j,k}(t,\phi)$ be defined as above satisfying 
\eqref{eq4.3}--\eqref{eq4.7} for $j=1,2$ and $k=0$. Then for 
\begin{equation}
U \equiv U_1(T,0) U^{\mathrm{aux}}_2(2T,0) U_1(T,0),\quad T\equiv 
\frac\pi{\eta\Omega} \sqrt N,
\end{equation}
we have
\begin{align}
\label{eq4.8}
U|0\rangle_1 |0\rangle_2|0\rangle_{\text{CM}} &= |0\rangle_1 |0\rangle_2 
|0\rangle_{\text{CM}},\\
\label{eq4.9}
U|1\rangle_1 |0\rangle_2 |0\rangle_{\text{CM}} &= |1\rangle_1 |0\rangle_2 
|0\rangle_{\text{CM}},\\
\label{eq4.10}
U|0\rangle_1 |1\rangle_2 |0\rangle_{\text{CM}} &= |0\rangle_1 |1\rangle_2 
|0\rangle_{\text{CM}},\\
\label{eq4.11}
U|1\rangle_1 |1\rangle_1 |0\rangle_{\text{CM}} &= -|1\rangle_1 |1\rangle_1 
|0\rangle_{\text{CM}}.
\end{align}
Consequently, by ignoring the last CM-bit $|0\rangle_{\text{CM}}$, we have the 
phase gate
\begin{equation}\label{eq4.11a}
U = Q_\pi,\quad \text{cf. (\ref{eq2.7}).}
\end{equation}
\end{thm}

\begin{proof}
We first verify \eqref{eq4.8}:
\begin{align}
U|0\rangle_1 |0\rangle_2 |0\rangle_{\text{CM}}
 =~&U_1(T,0) U^{\mathrm{aux}}_2 (2T,0) [U_1(T,0) |0\rangle_1 
|0\rangle_{\text{CM}}] |0\rangle_2\nonumber\\
=~&U_1(T,0) U^{\mathrm{aux}}_2(2T,0) [|0\rangle_1 |0\rangle_{\text{CM}}] 
|0\rangle_2\nonumber\\
&\quad \text{(by \eqref{eq4.4} because $|0\rangle_1 |0\rangle_{\text{CM}} \in 
\widetilde V_j$ for $j=1$)}\nonumber\\
=~&U_1(T,0) [U^{\mathrm{aux}}_2 (2T,0) |0\rangle_2 |0\rangle_{\text{CM}}] 
|0\rangle_1\nonumber\\
=~&U_1(T,0) [|0\rangle_2 |0\rangle_{\text{CM}}] |0\rangle_1\nonumber\\
&\quad \text{(by \eqref{eq4.7} because $|0\rangle_2 |0\rangle_{\text{CM}} \in 
\widetilde V^{\mathrm{aux}}_j$ for $j=2$)}\nonumber\\
=~&[U_1(T,0) |0\rangle_1 |0\rangle_{\text{CM}}] |0\rangle_2\nonumber\\
=~&|0\rangle_1 |0\rangle_{\text{CM}} |0\rangle_2\nonumber\\
&\quad \text{(by
\eqref{eq4.4} because $|0\rangle_2 |0\rangle_{\text{CM}} \in \widetilde 
V_j$ for $j=1$)}\nonumber\\
=~&|0\rangle_1 |0\rangle_2 |0\rangle_{\text{CM}}.
\end{align}
Next, we verify \eqref{eq4.9}:
\begin{align}
U|1\rangle_1 |0\rangle_2 |0\rangle_{\text{CM}}
=~&U_1(T,0) U^{\mathrm{aux}}_2 (2T,0) [U_1(T,0)
|1\rangle_0 |0\rangle_{\text{CM}}] |0\rangle_2\nonumber\\
%\label{eq4.12}
=~&U_1(T,0) U^{\mathrm{aux}}_2 (2T,0) [-i|0\rangle_1 |1\rangle_{\text{CM}}] 
|0\rangle_2\nonumber\\
&\quad \text{(using \eqref{eq4.3} with $k=0, \phi=0$ and $\cl E_kT=\pi/2$)}
\nonumber\\
=~&U_1(T,0) \{(-i) [U_2(2T,0) |0\rangle_2 |1\rangle_{\text{CM}}] 
|0\rangle_1\}\nonumber\\
=~&U_1(T,0) \{[i|0\rangle_2 |1\rangle_{\text{CM}}] |0\rangle_1\}\nonumber\\
&\quad \text{(using \eqref{eq4.6} and \eqref{eq4.3} with $k=0,\phi=0$ and 
$2\cl E_kT=\pi$)}\nonumber\\
=~&i[U_1(T,0)|0\rangle_1 |1\rangle_{\text{CM}}]|0\rangle_2\nonumber\\
%\label{eq4.13}
=~&i[(-i) |1\rangle_1 |0\rangle_{\text{CM}}] |0\rangle_2\nonumber\\
&\quad \text{(using \eqref{eq4.3} with $k=0,\phi=0$ and $\cl E_kT=\pi/2$)}
\nonumber\\
=~&|1\rangle_1 |0\rangle_2 |0\rangle_{\text{CM}}.
\end{align}
The verification of \eqref{eq4.10} can be done in the same way as that for 
\eqref{eq4.8}.

Finally, we verify \eqref{eq4.11}:
\begin{align}
U|1\rangle_1 |1\rangle_2 |0\rangle_{\text{CM}}
=~&U_1(T,0)U^{\mathrm{aux}}_2 (2T,0) [U_1(T,0) 
|1\rangle_1 |0\rangle_{\text{CM}}] 
|1\rangle_2\nonumber\\
=~&U_1(T,0) U^{\mathrm{aux}}_2(2T,0) [-i|0\rangle_1 |1\rangle_{\text{CM}}] 
|1\rangle_2\nonumber\\
&\quad  \text{(using \eqref{eq4.3} with $j=1, k=0, \phi=0$ and $\cl 
E_kT=\pi/2$)}\nonumber\\
=~&U_1(T,0) [(-i) U^{\mathrm{aux}}_2 (2T,0) |1\rangle_2 |1\rangle_{\text{CM}}] 
|0\rangle_1\nonumber\\
=~&U_1(T,0) [(-i) |1\rangle_2 |1\rangle_{\text{CM}}] |0\rangle_1\nonumber\\
&\quad \text{(by \eqref{eq4.7} because $|1\rangle_2 |1\rangle_{\text{CM}}\in 
\widetilde V^{\mathrm{aux}}_j$ for $j=2$)}\nonumber\\
=~&(-i) [U_1(T,0) |0\rangle_1 |1\rangle_{\text{CM}}] |1\rangle_2\nonumber\\
=~&(-i)[(-i) |1\rangle_1 |0\rangle_{\text{CM}}] |1\rangle_2\nonumber\\
&\quad \text{(using \eqref{eq4.3} with $j=1, k=0,\phi=0$ and $\cl 
E_kT=\pi/2$)}\nonumber\\
=~&-|1\rangle_1 |1\rangle_2 |0\rangle_{\text{CM}}.
\end{align}
The verifications are complete.
\end{proof}

The CNOT gate can now be obtained according to Corollary \ref{cor2.3}. As a 
consequence, we have the following.

\begin{thm}\label{thm4.4}
The quantum computer made of confined ions in a trap is universal.
\end{thm}

\begin{proof}
Use \eqref{eq2.8} and \eqref{eq4.11a} to deduce Corollary \ref{cor2.3}.
\end{proof}

A very similar gate has been implemented experimentally using a cold, confined 
Be$^+$ ion \cite{MMKIW}. To conclude this section, we mention two more schemes 
using ion traps.

\bigskip\noindent\textbf{(a)}\ The S\o{}rensen--M\o{}lmer scheme \cite{SM}.\\
Experimental inconveniences in implementation of the Cirac--Zoller scheme on 
confined ions include the requirement that the ions be initialized in the 
ground 
vibrational state $|0\rangle_{\text{CM}}$, which requires sophisticated laser 
cooling techniques, 
and the experimental observation that the CM vibration of a 
linear string of ions in a trap has limited coherence. 
The S\o{}rensen--M\o{}lmer 
scheme applies to a string of ions, but the ions can have thermal motion, as 
long 
as they remain within the Lamb--Dicke regime. The ions never populate a 
vibrational mode, so decoherence effects are reduced. The idea is to 
simultaneously excite an ion {\em pair\/} using two laser beams detuned by 
frequency $\delta$ from the upper and lower vibrational sidebands $\omega_0 + 
\omega_{\text{CM}} -\delta$ and $\omega_0 - \omega_{\text{CM}} + \delta$, so 
that the energy (or frequency) sum is $2\omega_0$, corresponding to excitation 
of both ions by two interfering transition paths 
from $|0,0\rangle|n\rangle$ to 
$|1,1\rangle|n\rangle$, without changing the vibrational state. The lasers are 
detuned from the red and blue sideband transitions of {\em single\/} ion 
excitation, so the vibration is only virtually (negligibly) excited, and the 
vibration levels are never actually populated. One laser beam is applied to 
each 
ion. For any two ions, each illuminated by {\em both\/} of the laser beams, an 
additional two interference paths exist, and the effective Rabi frequency for 
the two-ion transition becomes $\Omega_{\text{SM}} = 
-(\eta\Omega)^2/(\omega_{\text{CM}}-\delta)$, where $\Omega$ represents the
single ion Rabi frequency at the same laser power. The two-ion interaction 
Hamiltonian can be written 
$(\hbar\Omega_{\text{SM}}/2)(|0,0\rangle \langle1,1| 
+ 
|1,1\rangle \langle0,0|+$ $|1,0\rangle\langle 0,1| + 
|0,1\rangle\langle1,0|)$. The Rabi frequency can be approximated as 
$\Omega_{\text{SM}} = \eta^2\Omega^2/\delta$, since $\omega_{\text{CM}} - 
\delta\approx \delta$. Full entanglement is achieved at a particular 
interaction 
time $T=\pi/2\Omega_{\text{SM}}$, when the wavefunction becomes a coherent 
superposition of maximally entangled $|0,0\rangle$ and $|1,1\rangle$ states 
$|\psi\rangle = 2^{-1/2}(|0,0\rangle -i|1,1\rangle)$. Once entanglement is 
achieved, the CNOT gate can be implemented with additional single particle 
rotations, carried out analogously to those in the Cirac--Zoller scheme, with 
individual laser beams exciting single ions. Simultaneous entanglement of any 
{\em even\/} number of ions can be carried out similarly, although the CNOT 
gate 
applied to {\em any pair\/} of ions is sufficient for computation. The 
technique 
has been experimentally demonstrated with four ions in a small linear ion trap 
\cite{Sa}.

\bigskip\noindent\textbf{(b)}~The Jonathan--Plenio scheme \cite{JP}.\\
When an atomic transition is driven by a laser beam at resonance, an AC Stark 
shift (or light shift) results, changing the energies of the coupled 
levels due 
to the coupling to the laser field. The laser drive on an ion in a linear 
string 
in a trap produces Rabi transitions between the coupled levels, 
but if the Rabi 
frequency matches one of the CM vibration modes $\nu_p$ of the ion string, an 
exchange of excitation energy between the internal and vibrational variables 
also occurs. This exchange can be used to derive a CNOT gate between any two 
ions in the linear chain, based on an interference analogous to that of the 
S\o{}rensen--M\o{}lmer scheme discussed above. This is expected to result in 
relatively fast quantum  gates, but requires cold ions. 
When the light shift is 
used to drive virtual two-phonon transitions simultaneously, using more than 
one vibrational mode, it is predicted that gate speeds will be somewhat 
reduced, 
but the ions remain insensitive to heating, possibly offering 
an advantage over 
the S\o{}rensen--M\o{}lmer scheme. Two ion states of the form 
$|1,0\rangle|n_p\rangle$ and $|0,1\rangle|n_p\rangle$ are coupled. The 
effective time-independent Hamiltonian is written $-\hbar\omega[|1,0\rangle 
\langle 0,1| + |0,1\rangle\langle 1,0|]$ plus a sum over states $p$ which 
depend 
on the ion motion, but which can be cancelled by reversing the sign of a 
parameter half way through the operation. The states in the Hamiltonian are 
assumed now to be ``dressed'', i.e., formed from the original states coupled 
and 
shifted by interaction with the external field. The frequency is $\omega = 
(\Omega^2/2)\Sigma^N_{p=1} \eta_{1p} \eta_{2p} \nu_p/(\Omega^2-\nu^2_p)$. 
Maximally entangled states can be produced at the time $T = |\pi/4\omega|$. 
This 
proposed technique has not yet been experimentally demonstrated.

\section{Quantum Dots}\label{sec5}

Quantum dots are fabricated from semiconductor materials, metals, or small 
molecules. They work by confining electric charge quanta (i.e., spins)  in 
three 
dimensional boxes with electrostatic potentials. 
The spin of a charge quantum in a 
single quantum dot can be manipulated, i.e., single qubit operations, by 
applying pulsed local electromagnetic fields, through a scanning-probe tip, 
for 
example. Two-qubit operations can be achieved by spectroscopic manipulation or 
by a purely electrical gating of the tunneling barrier between neighboring 
quantum dots.

The spin-1/2 operator of an electron is given by
\begin{equation}\label{eq5.1}
\pmb{S} = (\sigma_x, \sigma_y,\sigma_z)^T = \sigma_x\pmb{e}_x + \sigma_y \pmb{
e}_y + \sigma_z \pmb{e}_z
\end{equation}
where $\sigma_x, \sigma_y$ and $\sigma_z$ are the usual Pauli matrices:
\begin{equation}\label{eq5.1a}
\sigma_x = \left[\begin{matrix} 0&1\\ 1&0\end{matrix}\right],\quad \sigma_y = 
\left[\begin{matrix} 0&-i\\ i&0\end{matrix}\right],\quad \sigma_z = 
\left[\begin{matrix} 1&0\\ 0&-1\end{matrix}\right],
\end{equation}
and 
\begin{equation}
\pmb{e}_x = \left[\begin{matrix} 1\\ 0\\ 0\end{matrix}\right],\quad 
\pmb{e}_y = 
\left[\begin{matrix} 0\\ 1\\ 0\end{matrix}\right],\quad \pmb{e}_z = 
\left[\begin{matrix} 0\\ 0\\ 1\end{matrix}\right]
\end{equation}
are the unit vectors in the directions of $x,y$ and $z$. Let $\pmb{S}_i$ and 
$\pmb{S}_j$ denote, respectively, the spin of the electric charge quanta at, 
respectively, the $i$-th and $j$-th location of the quantum dots. Then the 
usual 
physics of the Hubbard model \cite{AM} gives the Hamiltonian of the system of 
$n$ 
quantum dots as
\begin{equation}\label{eq5.2}
H = \sum^n_{j=1} \mu_B g_j(t) \pmb{B}_j(t) \cdot \pmb{S}_j + \sum_{1\le j<k\le 
n} J_{jk}(t) \pmb{S}_j\cdot \pmb{S}_k,
\end{equation}
where the first summation denotes the sum of energy due to the application of 
a magnetic field $\pmb{B}_j$ to the electron spin at dot $j$, while the second 
denotes the interaction Hamiltonian through the tunneling effect of a gate 
voltage 
applied between the dots, and
\begin{equation}
  \parbox[b]{0.9\textwidth}{%
\begin{itemize}
\item[$\mu_B$:] is the Bohr magneton;
\item[$g_j(t)$:] is the effective $g$-factor;
\item[$\pmb{B}_j(t)$:] is the applied magnetic field;
\item[$J_{jk}(t)$:] the time-dependent exchange constant \cite[see 
$\lbrack$10$\rbrack$ in 
the References therein]{LD}, with $J_{jk}(t) 
= 4t^2_{jk}(t)/u$, which is produced by the turning 
on and off of the tunneling 
matrix element $t_{ij}(t)$ between quantum dots $i$ and $j$, with $u$ 
being the charging energy of a single dot.
\end{itemize}}
\end{equation}
Note that for
\begin{equation}
\pmb{S}_j = \sigma^{(j)}_x \pmb{e}_x + \sigma^{(j)}_y \pmb{e}_y + 
\sigma^{(j)}_z 
\pmb{e}_z,\qquad j=1,2,\ldots, n,
\end{equation}
and
\begin{equation}
\pmb{B}_j(t) = b^{(j)}_x(t)\pmb{e}_x + b^{(j)}_y(t) \pmb{e}_y + b^{(j)}_z(t) 
\pmb{e}_z,\qquad j=1,2,\ldots, n,
\end{equation}
the dot products in \eqref{eq5.2} are defined by
\begin{align}\nonumber
\pmb{S}_j\cdot \pmb{S}_k &= \sigma^{(j)}_x \sigma^{(k)}_x + \sigma^{(j)}_y 
\sigma^{(k)}_y + \sigma^{(j)}_z \sigma^{(k)}_z,\\
\pmb{B}_j(t)\cdot\pmb{S}_j &= b^{(j)}_x(t) \sigma^{(j)}_x + b^{(j)}_y(t) 
\sigma^{(j)}_y + b^{(j)}_z(t)\sigma^{(j)}_z.
\end{align}

\begin{figure}
\centerline{\epsfig{file=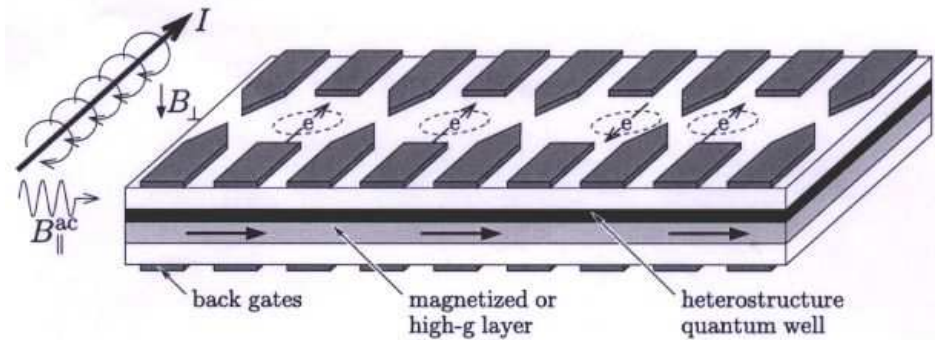,width=\textwidth}}
\caption{\label{fig5.1}%
Quantum dot array, controlled by electrical gating. The 
electrodes 
(dark gray) define quantum dots (circles) by confining electrons. The spin 1/2 
ground state (arrow) of the dot represents the qubit. These electrons can be 
moved by electrical gating into the magnetized or high-$g$ layer, producing 
locally different Zeeman splittings. Alternatively, magnetic field gradients 
can 
be applied, as e.g.\ produced by a current wire (indicated on the left of the 
dot-array). Then, since every dot-spin is subjected to a different Zeeman 
splitting, the spins can be addressed individually, e.g.\ through ESR 
(electron spin resonance) pulses of 
an additional in-plane magnetic ac field with the corresponding Larmor 
frequency 
$\omega_L = q\mu_B B_\bot/\hbar$. (In the figure, $B_\bot$ denotes the 
component of the magnetic field perpendicular to the array plane, while 
$B_{|~|}$ denotes that parallel to the plane.) Such mechanisms can be used for 
single-spin 
rotations and the initialization step. The exchange coupling between the 
quantum 
dots can be controlled by lowering the tunnel barrier between the dots. 
In this 
figure, the two rightmost dots are drawn schematically as tunnel-coupled. Such 
an exchange mechanism  can be used for the CNOT-gate operation involving two 
nearest neighbor qubits. The CNOT operator between distant qubits is achieved 
by 
swapping (via exchange) the qubits  first to a nearest neighbor position. The 
read-out of the spin state can be achieved via spin-dependent tunneling 
and SET devices \cite{LD}, or via a transport current passing the dot
\cite{LD}. 
Note that all 
spin operations, single and two spin operations, and spin  read-out, are 
controlled electrically via the charge of the electron and not 
via the magnetic 
moment of the spin. Thus, no control of local magnetic fields is required, and 
the spin is only used for storing the information. This spin-to-charge 
conversion is based on the Pauli principle and Coulomb interaction and allows 
for very fast switching times (typically picoseconds). A further advantage of 
this scheme is its scalability into an array of arbitrary size. (The figure is 
excerpted from Burkard, Engel and Loss \cite{BEL}, courtesy of Springer-Verlag 
Berlin Heidelberg, while the caption is excerpted from 
Schliemann and Loss \cite{SL}.)}
\end{figure}

In Fig.~\ref{fig5.1}, we include a quantum dot array design given in 
Burkard, Engel and Loss \cite{BEL}.

Again, by the universal quantum computing theorems in \S 2, we need only 
discuss the case of a system with two quantum dots. The spin-up $\uparrow$ and 
spin-down $\downarrow$ of the electric charge quanta in each dot are denoted
as $|0\rangle$ and $|1\rangle$, respectively. 
Then the underlying Hilbert space is
\begin{equation}\label{eq5.3a}
\cl H = \text{span}\{|00\rangle, |01\rangle, |10\rangle, |11\rangle\}.
\end{equation}
The wave function $|\psi(t)\rangle$ of the system has four components:
\begin{equation}\label{eq5.3b}
|\psi(t)\rangle = \psi_1(t)|00\rangle + \psi_2(t)|01\rangle + \psi_3(t) 
|10\rangle + \psi_4(t)|11\rangle,
\end{equation}
satisfying the Schr\"odinger equation
\begin{equation}\label{eq5.3}
\left\{\begin{array}{ll}\ds
i\hbar \frac\partial{\partial t} |\psi(t)\rangle = H(t) |\psi(t)\rangle,& 
t>0,\\[1.5ex]
|\psi(0)\rangle = |\psi_0\rangle \in \cl H,
\end{array}\right.
\end{equation}
where
\begin{equation}\label{eq5.4}
H(t) \equiv \frac\hbar2 [\pmb{\Omega}_1(t) \cdot \pmb{\sigma} + \pmb{
\Omega}_2(t) 
\cdot 
\pmb{\tau} + \omega(t) \pmb{\sigma}\cdot \pmb{\tau}],
\end{equation}
following from \eqref{eq5.2} by rewriting the notation
\begin{equation}\label{eq5.5}
\pmb{S}_1 =\pmb{\sigma},\quad \pmb{S}_2 = \pmb{\tau}\,; \quad \mu_B g_j(t) 
\pmb{
B}_j(t) = \frac\hbar2 \pmb{\Omega}_j(t),\quad j=1,2;\quad J_{12}(t) = 
\frac\hbar2 
\omega(t).
\end{equation}
The $\pmb{\Omega}_1(t), \pmb{\Omega}_2(t)$ and $\omega(t)$ 
in \eqref{eq5.4} are 
the {\em control pulses}.

Let us now give a brief heuristic physics discussion for the derivation of the 
Hamiltonian in \eqref{eq5.2} and \eqref{eq5.4} as a remark below.\medskip

\begin{rem}\label{rem4.2}
The suggestion to use quantum dots as qubits for the purposes of quantum
information processing in general, and quantum computing in particular, is
based on the following observations.
To set the stage, we note that a quantum dot is formed by a single
electron that is tightly bound to a ``center'' -- some distinguished small
area in a solid. 
Typically, the binding is strongly confining in one spatial direction (the
$z$ direction) while the electron has some freedom of motion in the plane
perpendicular to it (the $xy$ plane), with its dynamics governed by the forces
that bind it to its center.
In addition, there is the possibility of applying electric and magnetic
fields of varying strength and direction, whereby the experimenter can exert
some external control over the evolution of the electron's state.

In particular, a magnetic field $\brel{B}(t)$ will couple to the magnetic
moment of the electron, which gives rise to a term in the Hamilton operator
that is proportional to $\brel{B}\cdot\brel\sigma$, where $\brel\sigma$ is the
Pauli spin vector operator of the electron, so that
\begin{align}\label{eq5.8d}
  H=\frac12\hbar\brel\Omega(t)\cdot\brel\sigma
\end{align}
is the structure of the effective single-qubit Hamilton operator, with
$\brel\Omega(t)\propto\brel{B}(t)$ 
being externally controlled by the experimenter.
By suitably choosing $\brel{B}(t)$ various single-qubit gates can be
realized.

To suppress the unwanted effect of the Lorentz force (which couples to the
electron's charge, results in a time-dependent energy, and thus produces an
accumulated phase change), it is expedient to have the magnetic field vector
in the $xy$ plane, so that $\brel{B}\cdot\brel{e}_z=0$ is imposed. 
Nevertheless, general single qubit gates can be realized by a succession of
rotations of $\brel\sigma$ about axes in the $xy$ plane. 
This is just an application of Euler's classical result that general
rotations can be composed of successive rotations about two orthogonal axes
only (standard choice is the $z$ axis and the $x$ axis but, of course, the
$x$ axis and the $y$ axis serve this purpose just as well).

More complicated is the implementation of two-qubit gates.
Here one needs two quantum dots in close proximity, by which is meant that
they have a considerable interaction, predominantly originating in the
Coulomb repulsion of the charges of the two electrons. 
It is important that external electric fields can be used to modify the
effective interaction strength in a wide range, and in particular one can
``turn it off'' by separating the two quantum dots (or rather shielding them
from each other). 

Consider, thus, two quantum dots with some interaction potential in
addition to the potentials that bind them to their respective centers.
This situation is reminiscent of the hydrogen dimer, the $\mathrm{H}_2$
molecule, except that the confinement to a plane and the form of the binding
potential, and also of the effective interaction, are quite different. 
It is, indeed, not so simple to model the various potentials reasonably well,
and one must be practiced in the art of solid-state theory to do it well.
We shall, therefore, refer the interested reader to the specialized
literature, perhaps starting with  
\cite{BLD} and the references therein.

The general picture, however, can be grasped without getting involved with
such details. 
First, we note that at sufficiently low temperatures, only the ground state
and the first excited state of the interacting two-dot system will be
dynamically relevant, at least as long as we carefully avoid exciting other
states. 
For symmetry reasons, the ground state is a spin-singlet (total spin angular
momentum of $0\hbar$) and has a symmetric spatial wave function, whereas 
the first excited state is a spin-triplet 
(total spin angular momentum of $1\hbar$)
and has an antisymmetric spatial wave function. 
The excited state is long-lived because
triplet-to-singlet transitions tend to have very small matrix elements.

The total spin angular momentum vector operator 
is $\brel{S}=\frac12\hbar(\brel\sigma+\brel\tau)$ 
where we denote
the Pauli vector operators of the two electrons 
by $\brel\sigma$ and $\brel\tau$, respectively.
The eigenvalues of $\brel{S}^2$ are $0\hbar^2$ in the spin singlet and
$2\hbar^2$ in the spin triplet. In view of $\brel\sigma^2=3\pmb{1}$ and
$\brel\tau^2=3\pmb{1}$, this says that $\brel\sigma\cdot\brel\tau$ has 
eigenvalue
$-3$ in the singlet and $+1$ in the triplet.
As a consequence, 
\begin{equation}
\left\{\begin{array}{l}
\dfrac14\bigl(\pmb{1}-\brel\sigma\cdot\brel\tau\bigr)
\text{ projects on the singlet, and}\\
\noalign{\smallskip}
\dfrac14\bigl(3\pmb{1}+\brel\sigma\cdot\brel\tau\bigr)
\text{ projects on the triplet.}\end{array}\right.
\end{equation}
Effectively, then, they project on the ground state and the excited state,
respectively. 
Denoting by $E_0$ the ground state energy, and by $E_1=E_0+2\hbar\omega$ 
that of the excited state, we have
\begin{align}\nonumber
  H&=\frac14\bigl(\pmb{1}-\brel\sigma\cdot\brel\tau\bigr)E_0
   +\frac14\bigl(3\pmb{1}+\brel\sigma\cdot\brel\tau\bigr)E_1\\   
&=\frac14\bigl(E_0+3E_1)\pmb{1}+\frac14(E_1-E_0)\brel\sigma\cdot\brel\tau
\end{align}
or, after dropping the irrelevant additive constant,
\begin{align}\label{eq5.8g}
  H=\frac12\hbar\omega(t)\brel\sigma\cdot\brel\tau
\end{align}
for the effective Hamilton operator of the coupling between the two quantum
dots.
We have indicated the externally controlled time dependence of $\omega$, the
``turning on'' and ``turning off'' of the interaction at the experimenter's
discretion, by $t$.

In summary, then, we have two terms of the form (\ref{eq5.8d}), one for
each quantum dot, and the interaction term (\ref{eq5.8g}) in the effective
Hamilton operator
\begin{equation}
  H=\frac12\hbar\bigl[\brel\Omega_1(t)\cdot\brel\sigma
+\brel\Omega_2(t)\cdot\brel\tau
+\omega(t)\brel\sigma\cdot\brel\tau\bigr]\,.
\end{equation}
Within reasonable ranges, the experimenter is capable of realizing any
$\brel\Omega_1(t)$, $\brel\Omega_2(t)$, and $\omega(t)$, 
where it is fully sufficient to have two of them vanishing at any instant.
$\hfill\square$
\end{rem}
\medskip

\begin{lem}\label{lem5.1}
Let $\pmb{\sigma}$ and $\pmb{\tau}$ be defined as in \eqref{eq5.5}. Then we 
have
\begin{align}\label{eq5.9a}
&{\rm ~(i)}~~\left(\dfrac{1+\pmb{\sigma}\cdot\pmb{\tau}}{2}\right)^2 = 
\pmb{1} \text{ and }
(\pmb{\sigma}\cdot\pmb{\tau})^2 = 3\pmb{1} - 
2\pmb{\sigma} \cdot 
\pmb{\tau}.\\
\label{eq5.10}
&{\rm (ii)}~~U_{\mathrm{sw}} = \frac12 (\pmb{1} 
+\pmb{\sigma}\cdot 
\pmb{\tau}) = 
U^{\dagger}_{\mathrm{sw}} = U^{-1}_{\mathrm{sw}}, 
\text{ where $U_{\mathrm{sw}}$ is the swapping gate}
\end{align}
\begin{equation}
U_{\mathrm{sw}}|jk\rangle 
= |kj\rangle\quad \text{for}\quad j,k=0,1; \text{cf.\ 
(\ref{eq2.5}).}
\end{equation}

\hspace{.3in}{\rm (iii)}
\begin{equation}\label{eq5.9}
U^{\dagger}_{\mathrm{sw}} \pmb{\sigma} U_{\mathrm{sw}} 
= \pmb{\tau},\quad U^{\dagger}_{\mathrm{sw}} \pmb{\tau} U_{\mathrm{sw}} = 
\pmb{\tau}
\end{equation}
\begin{equation}\label{eq5.12a}
\hspace{-3.1in}{\rm (iv)}~~U^2_{\mathrm{sw}}  = \pmb{1}.
\end{equation}
\end{lem}

\begin{proof}
Let us recall first how the various $\sigma_j$ and $\tau_j$
transform the basis vectors, for $j=x,y,z$:
\begin{align}
  \sigma_x|0\cdot\rangle&=|1\cdot\rangle\,,\qquad
  &\sigma_x|1\cdot\rangle&=|0\cdot\rangle\,,\nonumber\\
  \sigma_y|0\cdot\rangle&=i|1\cdot\rangle\,,\qquad
  &\sigma_y|1\cdot\rangle&=-i|0\cdot\rangle\,,\nonumber\\
  \sigma_z|0\cdot\rangle&=|0\cdot\rangle\,,\qquad
  &\sigma_z|1\cdot\rangle&=-|1\cdot\rangle\,,
\end{align}
and likewise for $\tau_x$, $\tau_y$, and $\tau_z$ 
acting on $|\cdot0\rangle$ and $|\cdot1\rangle$.
Thus
\begin{align}
  \brel\sigma\cdot\brel\tau|00\rangle
&=(1^2+i^2)|11\rangle+|00\rangle=|00\rangle\,,\nonumber\\
  \brel\sigma\cdot\brel\tau|01\rangle
&=(1^2-i^2)|10\rangle-|01\rangle=2|10\rangle-|01\rangle\,,\nonumber\\
  \brel\sigma\cdot\brel\tau|10\rangle
&=(1^2-i^2)|01\rangle+|10\rangle=2|01\rangle-|10\rangle\,,\nonumber\\
  \brel\sigma\cdot\brel\tau|11\rangle
&=(1^2+i^2)|00\rangle+|11\rangle=|11\rangle\,,
\end{align}
and we see that the $4\times4$ matrix for $\brel\sigma\cdot\brel\tau$ is given 
by
\begin{align}
  \brel\sigma\cdot\brel\tau=\left[
    \begin{matrix}
      1&0&0&0\\ 0&-1&2&0\\ 0&2&-1&0 \\ 0&0&0&1
    \end{matrix}\right]\,,
\end{align}
and the first equation in \eqref{eq5.9a} follows.

We read off that $\brel\sigma\cdot\brel\tau$ has a 3-fold eigenvalue $+1$ and
a single eigenvalue $-3$.
The respective projectors to the subspaces characterized by these eigenvalues
are easily verified to be
\begin{align}
\bb P_1 \equiv  
\frac{1}{4}(3\pmb{1}+\brel\sigma\cdot\brel\tau)\quad\mbox{and}\quad
\bb P_2\equiv  \frac{1}{4}(\pmb{1}-\brel\sigma\cdot\brel\tau)\,,
\end{align}
where $\bb P^2_j = \bb P_j$ for $j=1,2$, and $\bb P_j\bb P_k=0$ for $j\ne k$.
Therefore, for any sufficiently well-behaved function, including polynomials  
and analytic functions, of
$\brel\sigma\cdot\brel\tau$, 
\begin{align}\nonumber
  f(\brel\sigma\cdot\brel\tau)&=
f(1)\frac{1}{4}(3\pmb{1}+\brel\sigma\cdot\brel\tau)
+f(-3) \frac{1}{4}(\pmb{1}-\brel\sigma\cdot\brel\tau)
\\  \label{eq5.10a}
&=\frac{1}{4}\left[3f(1)+f(-3)\right]\pmb{1}
+\frac{1}{4}\left[f(1)-f(-3)\right]\brel\sigma\cdot\brel\tau\,,
\end{align}
according to the spectral theorem.
As a first application, consider $f(x)=x^2$ and find
\begin{align}
  (\brel\sigma\cdot\brel\tau)^2=3\pmb{1}-2\brel\sigma\cdot\brel\tau\,,
\end{align}
which is the second equation in \eqref{eq5.9a}.

Further, in view of the matrix for $U_{\mathrm{sw}}$,
\begin{align}
  U_{\mathrm{sw}}=\frac{1}{2}(\pmb{1}+\brel\sigma\cdot\brel\tau)
=\left[\begin{matrix}
      1&0&0&0\\ 0&0&1&0\\ 0&1&0&0 \\ 0&0&0&1
    \end{matrix}\right]
\end{align}
the swapping gate really swaps: $U_{\mathrm{sw}}|jk\rangle=|kj\rangle$ (no
effect on $|00\rangle$ and $|11\rangle$, whereas $|01\rangle$ and
$|10\rangle$ are interchanged).
To verify
\begin{align}
  U_{\mathrm{sw}}\brel\sigma=\brel\tau U_{\mathrm{sw}}
\end{align}
it should suffice to inspect the pair
\begin{align}
  \sigma_x&=|00\rangle\langle10|+|01\rangle\langle11|+
           |10\rangle\langle00|+|11\rangle\langle01|\\
\intertext{and}
    \tau_x&=|00\rangle\langle01|+|10\rangle\langle11|+
           |01\rangle\langle00|+|11\rangle\langle10|\,,
\end{align}
for example. 
One can also have a purely algebraic, but more lengthy, argument
that exploits nothing but the basic identities, such as $\sigma_x^2=\pmb{1}$
and $\sigma_1\sigma_2=i\sigma_3$, etc. Thus, the rest easily follows.
\end{proof}

\begin{thm}\label{thm5.2}
Denote by $U(t)$  the time evolution operator for the quantum system 
\eqref{eq5.3} and \eqref{eq5.4} for time duration $t\in [0,T]$. Choose $\pmb{
\Omega}_1(t) = \pmb{\Omega}_2(t) = 0$ in \eqref{eq5.4} and let $\omega(t)$ 
therein satisfies
\begin{equation}\label{eq5.11}
\int^T_0 \omega(t)dt = \frac\pi2.
\end{equation}
Then we have $U(T) =-e^{\pi i/4} U_{\mathrm{sw}}$, i.e., 
$U(T)$ is the swapping gate 
(with a nonessential phase factor $-e^{\pi i/4}$.)
\end{thm}

\begin{proof}
By assumptions, we have now
\begin{equation}
H(t) = \omega(t) \pmb{\sigma}\cdot\pmb{\tau}/2.
\end{equation}
Since $\omega(t)$ is scalar-valued, we have the commutativity
\begin{equation}
H(t_1) H(t_2) =H(t_2)H(t_1),\quad \text{for any}\quad t_1,t_2\in [0,T].
\end{equation}
Therefore
\begin{align}
U(T) &= e^{-i \int^T_0 H(t)dt/\hbar} 
= e^{\left[-\frac{i}2 \int^T_0 \omega(t) 
dt\right]\pmb{\sigma}\cdot \pmb{\tau}}\nonumber\\
&= e^{-i\phi\pmb{\sigma}\cdot\pmb{\tau}}\hspace{1.5in} 
\left(\phi\equiv \frac12 
\int^T_0 \omega(t)dt\right)\nonumber\\
\label{eq5.12}
&= \cos(\phi\pmb{\sigma}\cdot\pmb{\tau}) - i \sin(\phi\pmb{\sigma}\cdot 
\pmb{\tau}),
\end{align}
where $e^{-i\phi\pmb{\sigma}\cdot\pmb{\tau}}$, $\cos(\phi\pmb{\sigma}\cdot 
\pmb{\tau})$ and $\sin(\phi\pmb{\sigma}\cdot\pmb{\tau})$ are $4\times 4$ 
matrices. 
We now utilize \eqref{eq5.10a} to calculate the exponential matrix $U(T)$:
\begin{equation}\label{eq5.13}
U(T) = e^{-i\phi\pmb{\sigma}\cdot\pmb{\tau}} 
= e^{-i\phi}\cdot \frac14 (3\pmb{1} + \pmb{\sigma} \cdot\pmb{\tau}) + 
e^{-3i\phi} \cdot \frac14 (\pmb{1} - \pmb{\sigma}\cdot\pmb{\tau}).
\end{equation}
Thus, with a little manipulation, \eqref{eq5.13} becomes
\begin{align}
U(T) &= e^{i\phi} \left[\cos(2\phi) \pmb{1} - i\sin(2\phi) \frac{\pmb{1} + 
\pmb{\sigma}\cdot\pmb{\tau}}2\right]\nonumber\\
\label{eq5.14}
&= e^{i\phi} [\cos(2\phi)\pmb{1} - i\sin(2\phi) U_{\mathrm{sw}}],\quad 
\text{by (\ref{eq5.10}).}
\end{align}
Choosing $\phi = \pi/4$, 
we obtain the desired conclusion.
\end{proof}

\begin{cor}\label{cor5.3}
The square roots of the swapping gate, $U^{1/2}_{\mathrm{sw}}$,  are
\begin{equation}\label{eq5.15}
U^{1/2}_{\mathrm{sw}} = \frac{e^{\pm\pi i/4}}{\sqrt 2} 
(\pmb{1} \mp iU_{\mathrm{sw}}).
\end{equation}
\end{cor}

\begin{proof}
From \eqref{eq5.14}, we first obtain
\begin{equation}
U_{\mathrm{sw}} = ie^{-\frac{\pi i}4} U(T).
\end{equation}
Then use $\phi = \pm \pi/8$ in \eqref{eq5.14} to obtain
\begin{equation}
U^{1/2}_{\mathrm{sw}} = (ie^{-\frac{\pi i}4})^{1/2} e^{\pm\pi i/8} 
\left[\frac1{\sqrt 2} 
(\pmb{1} \mp i U_{\mathrm{sw}})\right]
\end{equation}
and the desired conclusion. 
(Note that these two square roots of $U_{\mathrm{sw}}$ 
reflect the choices of $\sqrt 1=1$ and
\begin{equation}
\text{the square root of $-1=\pm i$}
\end{equation}
for the square roots of the eigenvalues of $U_{\mathrm{sw}}$.)
\end{proof}

Theorem \ref{thm5.2} gives us the choice of control pulses $\omega(t)$ (with 
$\pmb{\Omega}_1(t)$ and $\pmb{\Omega}_2(t)$ being set to 0) which yields the 
swapping gate $U_{\mathrm{sw}}$, which does not cause entanglement. However, 
$U^{1/2}_{\mathrm{sw}}$ in Corollary \ref{cor5.3} causes entanglement. 
See Corollary 
\ref{cor5.5} shortly.

\begin{thm}\label{thm5.4}
Let $\phi,\theta \in [0,2\pi]$ be given. Denote $\pmb{e}(\phi) = \cos\phi\pmb{
e}_x + \sin \phi \pmb{e}_y + 0 \pmb{e}_z$ for the given $\phi$. Let $U_{1,\pmb{
\Omega}_1}(t)$ be the time evolution operator corresponding to the quantum 
system 
\eqref{eq5.3} and \eqref{eq5.4} for $t\in [0,T]$ where the pulses are chosen 
such that
\begin{equation}\label{eq5.16}
\pmb{\Omega}_1(t) = \Omega_1(t) \pmb{e}(\phi),\quad 
\pmb{\Omega}_2(t) = 0,\quad 
\omega(t) = 0,\qquad t\in [0,T],
\end{equation}
with $\Omega_1(t)$ satisfying
\begin{equation}\label{eq5.17}
\int^T_0 \Omega_1(t) dt = 2\theta,\quad \text{for the given } \theta.
\end{equation}
Then the action of $U_{1,\pmb{\Omega}_1}(t)$ on the first qubit satisfies
\begin{equation}
U_{1,\pmb{\Omega}_1}(t) = U_{\theta,\phi}, \text{ the 1-bit unitary rotation 
gate 
(\ref{eq2.6}).}
\end{equation}
\end{thm}

\begin{proof}
We have
\begin{align}
U_{\theta,\phi} &= \left[\begin{matrix} \cos\theta&-ie^{-i\phi}\sin\theta\\ 
-ie^{i\phi}\sin\theta&\cos\theta\end{matrix}\right]\nonumber\\
&= \cos\theta \pmb{1}  -ie^{-i\phi} \sin\theta \left(\frac{\sigma_x - 
i\sigma_y}2\right) - ie^{i\phi} \sin\theta 
\left(\frac{\sigma_x-i\sigma_y}2\right)\nonumber\\
&= \cos\theta \pmb{1} - i \sin \theta \cos \phi \sigma_x - i\sin \theta \sin 
\phi \sigma_y\nonumber\\
&= \cos\theta \pmb{1} - i\sin\theta(\cos\phi \sigma_x 
+\sin\phi\sigma_y)\nonumber\\
&= \cos\theta \pmb{1} - i\sin \theta \pmb{e}(\phi) \cdot\pmb{\sigma}\nonumber\\
\label{eq5.18}
&= e^{-i\theta \pmb{e}(\phi)\cdot\pmb{\sigma}},
\end{align}
noting that in the above, we have utilized the fact that the $2\times 2$ matrix
\begin{equation}
\pmb{e}(\phi)\cdot \pmb{\sigma} = \left[\begin{matrix}
0&\cos\phi - i\sin\phi\\ 
\cos \phi+ i\sin \phi&0\end{matrix}\right]
\end{equation}
satisfies $(\pmb{e}(\phi)\cdot\pmb{\sigma})^{2n} = \pmb{1}$ for 
$n=0,1,2,\ldots$~.

With the choices of the pulses as given in \eqref{eq5.16}, we see that the 
second 
qubit remains steady in the time-evolution of the system. 
The Hamiltonian, now, is
\begin{equation}
H_1(t) = \frac\hbar2 \Omega_1(t) \pmb{e}_1(\phi)\cdot\pmb{\sigma}
\end{equation}
and acts only on the first  qubit (where the subscript 1 of $\pmb{e}_1(\phi)$ 
denotes that this is the vector $\pmb{e}(\phi)$ for the first bit). Because 
$\Omega_1(t)$ is scalar-valued, we 
have
\begin{equation}
H_1(t_1) H_1(t_2) = H_1(t_2) H_1(t_1)\quad \text{for any}\quad t_1,t_2\in 
[0,T].
\end{equation}
Thus
\begin{align}
U_{1,\pmb{\Omega}_1}(T)&= e^{-\frac{i}2 \int^T_0 \Omega_1(t) \pmb{e}_1(\phi) 
\cdot \pmb{\sigma} \ dt}\nonumber\\
&= e^{\left[-\frac{i}2 \int^T_0\Omega_1(t) dt\right] \pmb{e}_1(\phi)\cdot 
\pmb{\sigma}}\nonumber\\
&= e^{-i\theta\pmb{e}_1(\phi)\cdot \pmb{\sigma}},\hspace{1.3in} \text{(by 
\eqref{eq5.17})}
\end{align}
using \eqref{eq5.18}. The proof is complete.
\end{proof}

We may define $U_{2,\pmb{\Omega}_2}$ in a similar way as in Theorem 
\ref{thm5.4}.

\begin{cor}\label{cor5.5}
The quantum phase gate $Q_\pi$  \eqref{eq2.7} is given by
\begin{equation}
Q_\pi = (-i) U_{1,\pmb{\Omega}^{(2)}_1} 
U_{2,\pmb{\Omega}_2}
U^{1/2}_{\mathrm{sw}} U_{1,\pmb{\Omega}^{(1)}_1} U^{1/2}_{\mathrm{sw}},
\end{equation}
where 
\begin{equation}\label{eq5.21a}
\left\{\begin{array}{l}
\ds\int \pmb{\Omega}^{(1)}_1 (t)\, dt = -\pi \pmb{e}_{1z},\\
\ds\int \pmb{\Omega}^{(2)}_1 (t)\, dt = \pi \pmb{e}_{1z}/2,\\
\ds\int \pmb{\Omega}_2(t) \,dt = -\pi \pmb{e}_{2z}/2,
\end{array}\right.
\end{equation}
and $\pmb{e}_{1z}, \pmb{e}_{2z}$  denote the $\pmb{e}_z$ vector of, 
respectively, the first and the second qubit.
\end{cor}

\begin{rem}
In order to realize this succession of gates, only one of the
$\pmb{\Omega}(t)$ in \eqref{eq5.21a} is nonzero at any given instant $t$, with
the duration when $\pmb{\Omega}^{(1)}_1(t)\neq0$ earlier than that when
 $\pmb{\Omega}_2(t)\neq0$, and that when  $\pmb{\Omega}^{(2)}_1(t)\neq0$
even later. Earliest is the period when $\omega(t)\neq0$ for the first
$U^{1/2}_{\mathrm{sw}}$, and another  period when $\omega(t)\neq0$ is
intermediate between those when $\pmb{\Omega}^{(1)}_1(t)\neq0$ 
and $\pmb{\Omega}_2(t)\neq0$.
\end{rem}

\begin{proof}
Define
\begin{equation}\label{eq5.24.1}
U_{\text{XOR}} \equiv e^{\frac{\pi i}4 \sigma_z} e^{-\frac{\pi i}4 \tau_z} 
U^{1/2}_{\mathrm{sw}} e^{i\frac\pi2 \sigma_z} U^{1/2}_{\mathrm{sw}},
\end{equation}
with $U^{1/2}_{\mathrm{sw}} = \frac{e^{-\frac\pi4 i}}{\sqrt 2} 
(\pmb{1} + iU_{\mathrm{sw}})$ 
chosen from \eqref{eq5.15}. Then it is straightforward to check that
\begin{align}\nonumber
U_{\text{XOR}}|00\rangle &= |00\rangle (i), \quad U_{\text{XOR}} |01\rangle = 
|01\rangle(i),\\
U_{\text{XOR}}|10\rangle &= |10\rangle (i),\quad U_{\text{XOR}} |11\rangle  = 
|11\rangle (-i),
\end{align}
so that
\begin{align}
U_{\text{XOR}} &= i(|00\rangle \langle 00| + |01\rangle \langle 01| + 
|10\rangle\langle 10| - |11\rangle\langle11|)\nonumber\\
&= iQ_\pi.
\end{align}
\end{proof}
\medskip

\begin{rem}\label{rem4.2a}
In \cite[eq.~(2)]{LD}, Loss and DiVincenzo write their $U_{\text{XOR}}$ 
gate as
\begin{equation}
U_{\text{XOR}} = e^{i\frac\pi2 S^z_1} 
e^{-i\frac\pi2 S^z_2} U^{1/2}_{\mathrm{sw}} 
e^{i\pi 
S^z_1} U^{1/2}_{\mathrm{sw}}.
\end{equation}
Thus there is a difference in notation between the above and \eqref{eq5.24.1}:
\begin{equation}
S^2_1 = \sigma_z/2,\quad S^z_2 = \tau_z/2. 
\end{equation}
\hfill$\square$
\end{rem}
\medskip

\begin{thm}\label{thm5.6}
The quantum computer made of quantum dots is universal.
\end{thm}

\begin{proof}
This follows easily from \eqref{eq2.8}, Theorem \ref{thm5.4}, Corollary 
\ref{cor5.5}, and finally Corollary \ref{cor2.3}.
\end{proof}

Throughout this paper we have been employing the Schr\"odinger
point of view and could therefore afford to identify the ket vectors with
particular numerical column vectors, the bra vectors with the corresponding
row vectors, and we did not distinguish between operators and matrices.
Now, let us give a brief, heuristic exposition
from the point of view of the Heisenberg picture, in which it is important to
discriminate between kets and their columns, bras and their rows, operators
and their matrices. 
Dirac's classic text \cite{DiracQMbook} and the recent textbook by Schwinger
\cite{SchwingerQMbook} treat quantum mechanics consistently in the Heisenberg
picture. 

For a start we return to (\ref{eq5.3a}) and (\ref{eq5.3b}) where each 
$|jk\rangle$ is a joint 
eigenket of $\sigma_z$ and $\tau_z$ with eigenvalues $1-2j$ and $1-2k$, 
respectively for $j,k=0,1$.
A simultaneous measurement of $\sigma_z$ and $\tau_z$ yields the actual one of
the four pairs of values and tells us which of the four possibilities is the
case.
 
Now, a measurement of $\sigma_z$, say, at time $t_1$ is not the same as a
measurement of $\sigma_z$ at time $t_2$, one really has to perform two
different experiments (perhaps using the same lab equipment twice). 
Therefore, according to the reasoning of Heisenberg and Dirac, just to
mention the main quantum physicists, we should also have two mathematical 
symbols for
these two measurements: $\sigma_z(t_1)$ and  $\sigma_z(t_2)$.
As a consequence, then, also the eigenkets must refer to the time of
measurement: $|jk,t_1\rangle$ and $|jk,t_2\rangle$.
Accordingly, the eigenvector equations read
\begin{align}\label{eq19}
\setlength{\arraycolsep}{2pt}
  \left.\begin{array}{rcl}
  \sigma_z(t)|jk,t\rangle&=&|jk,t\rangle(1-2j)\\[1ex]
  \tau_z(t)|jk,t\rangle&=&|jk,t\rangle(1-2k)
\end{array}\right\}
\mbox{for all $t$, for $j,k=0,1$.}
\setlength{\arraycolsep}{5pt}
\end{align}
On the other hand, the state of the system, which is typically specified by a
statement of the kind ``at time $t_0$ both spins were up in the $z$
direction'' (formally: state ket $|\Psi\rangle=|00,t_0\rangle$), 
does not depend on time by its nature. 
So, rather than (\ref{eq5.3b}), we write
\begin{align}
  \label{eq5.20}
|\Psi\rangle 
= \psi_1(t)|00,t\rangle + \psi_2(t)|01,t\rangle + \psi_3(t) 
|10,t\rangle + \psi_4(t)|11,t\rangle\,,
\end{align}
where the \emph{dynamical} time dependences of the kets $|00,t\rangle$, \dots,
$|11,t\rangle$ is compensated for by
the \emph{parametric} time dependence of the probability amplitudes
$\psi_1(t)$, \dots, $\psi_4(t)$, so that the state ket
$|\Psi\rangle$ is constant in time, as it should be.

The probability amplitudes are thus given by
\begin{align} \label{eq5.21}
  \psi_1(t)=\langle00,t|\Psi\rangle\,,\quad
  \psi_2(t)=\langle01,t|\Psi\rangle\,,\quad
  \psi_3(t)=\langle10,t|\Psi\rangle\,,\quad
  \psi_4(t)=\langle11,t|\Psi\rangle\,,
\end{align}
and they have the same meaning as before: $|\psi_2(t)|^2$, for example, is
the probability that the first spin is found up and the second down at time
$t$. And, of course, they obey the same Schr\"odinger equation as before,
namely (\ref{eq5.3}) where 
$|\psi(t)\rangle$ is just the column
$\left[\psi_1(t)\ \psi_2(t)\ \psi_3(t)\ \psi_4(t)\right]^\mathrm{T}$
and $H(t)$ the $4\times4$ matrix that is meant in (\ref{eq5.4}).

With the Hamilton \emph{operator} (not matrix!)
\begin{align}\label{eq5.22}
  H\bigl(\brel\sigma(t),\brel\tau(t),t\bigr)=
\frac\hbar2 [\brel \Omega_1(t) \cdot \brel\sigma(t) 
+ \brel \Omega_2(t) \cdot \brel\tau(t) 
+ \omega(t) \brel\sigma(t)\cdot \brel\tau(t)]\,,
\end{align}
which has a dynamical time dependence in $\brel\sigma(t)$ and $\brel\tau(t)$ 
and a
parametric time dependence in $\brel\Omega_j(t)$, $j=1,2$, and $\omega(t)$,
the four rows of this Schr\"odinger equation are just
\begin{align}\label{eq5.23}
  i\hbar\frac{\partial}{\partial t}\langle jk,t|\Psi\rangle=
\langle jk,t| H\bigl(\brel\sigma(t),\brel\tau(t),t\bigr)|\Psi\rangle
\qquad\mbox{for $jk=00,01,10,11$}
\end{align}
and the elements of the $4\times4$ matrix of (\ref{eq5.4}) are
\begin{align}\label{eq5.24}
 \langle jk,t| H\bigl(\brel\sigma(t),\brel\tau(t),t\bigr)|lm,t\rangle 
\qquad\mbox{with $jk=00,01,10,11$ and $lm=00,01,10,11$.}
\end{align}

The Pauli matrices in (\ref{eq5.1a}) appear when we express the components
of $\brel\sigma(t)$ explicitly in terms of the eigenkets and eigenbras of
$\sigma_z(t)$ and $\tau_z(t)$, 
\begin{align}\label{eq5.25}
  \sigma_x(t)&=\sum_{k=0,1}\bigl(
    |0k,t\rangle\langle1k,t|+|1k,t\rangle\langle0k,t|\bigr)
\nonumber\\&=\sum_{k=0,1}
\left[\begin{matrix}|0k,t\rangle\\|1k,t\rangle\end{matrix}\right]
\left[\begin{matrix}0&1\\1&0\end{matrix}\right]
\bigl[\langle0k,t|,\langle1k,t|\bigr]\,,\nonumber\\
 \sigma_y(t)&=\sum_{k=0,1}
\left[\begin{matrix}|0k,t\rangle\\|1k,t\rangle\end{matrix}\right]
\left[\begin{matrix}0&-i\\i&0\end{matrix}\right]
\bigl[\langle0k,t|,\langle1k,t|\bigr]\,,\nonumber\\
 \sigma_z(t)&=\sum_{k=0,1}
\left[\begin{matrix}|0k,t\rangle\\|1k,t\rangle\end{matrix}\right]
\left[\begin{matrix}1&0\\0&-1\end{matrix}\right]
\bigl[\langle0k,t|,\langle1k,t|\bigr]\,,
\end{align}
and likewise for $\brel\tau(t)$, compactly written as
\begin{align}\label{eq5.26}
  \brel\tau(t)=\sum_{j=0,1}
\left[\begin{matrix}|j0,t\rangle\\|j1,t\rangle\end{matrix}\right]
\left(\left[\begin{matrix}0&1\\1&0\end{matrix}\right]\brel{e}_x
+\left[\begin{matrix}0&-i\\i&0\end{matrix}\right]\brel{e}_y
+\left[\begin{matrix}1&0\\0&-1\end{matrix}\right]\brel{e}_z\right)
\bigl[\langle j0,t|,\langle j1,t|\bigr]\,.
\end{align}
Note how numerical $2\times2$ matrices are sandwiched by 2-component columns
of kets and 2-component rows of bras to form linear combinations of products
of the dyadic ket-times-bra type.

The Schr\"odinger equation (\ref{eq5.23}) is the equation of motion of the
bras $\langle jk,t|$ and those of the kets $|jk,t\rangle$ is the adjoint
system. 
The equation of motion for the operator $\brel\sigma(t)$
follows from them,
\begin{align}\label{eq5.27}
  i\hbar\frac{d}{dt}\brel\sigma(t)
=\brel\sigma(t)H_t-H_t\brel\sigma(t)=\bigl[\brel\sigma(t),H_t\bigr]\,,
\end{align}
and likewise for $\brel\tau(t)$,
\begin{align}\label{eq5.28}
  i\hbar\frac{d}{dt}\brel\tau(t)
=\bigl[\brel\tau(t),H_t\bigr]\,,
\end{align}
where $H_t$ abbreviates $H\bigl(\brel\sigma(t),\brel\tau(t),t\bigr)$,
the Hamilton operator of (\ref{eq5.22}). 
These are particular examples of the more general \emph{Heisenberg equation
  of motion}, which states that any operator -- here: any function $O_t$ of
$\brel\sigma(t)$, $\brel\tau(t)$, and $t$ itself -- obeys
\begin{align}\label{eq5.29}
  \frac{d}{dt}O_t= \frac{\partial}{\partial t}O_t
  +\frac{1}{i\hbar}\bigl[O_t,H_t\bigr]\,,
\end{align}
where $\frac{d}{dt}$ is the total time derivative whereas
$\frac{\partial}{\partial t}$ differentiates only the parametric time
derivative, so that the commutator term accounts for the dynamical change in
time of $O_t$.
Note that (\ref{eq5.29}) has exactly the same structure as the Hamilton
equation of motion in classical mechanics.

A particular $O_t$ is the unitary evolution operator 
$U_{t_0,t}\equiv U\bigl(t_0;\brel\sigma(t),\brel\tau(t),t\bigr)$ 
that transforms a bra at time $t_0$ into the corresponding bra at time $t$ 
(earlier or later),
\begin{align}\label{eq5.30}
  \langle\dots,t|=\langle\dots,t_0|U_{t_0,t}\,,
\end{align}
where the ellipsis stands for any corresponding set of quantum numbers (such
as $00$, \dots $11$). 
$U_{t_0,t}$ has a parametric dependence on both $t_0$ and $t$, 
in addition to the dynamical $t$ dependence that originates in 
$\brel\sigma(t)$ and $\brel\tau(t)$. 
The group property $U_{t_0,t_1}U_{t_1,t_2}=U_{t_0,t_2}$ follows from the
definition and so does the unitarity, 
and thus
\begin{align}\label{eq5.31}
  {U_{t_0,t}}^\dagger={U_{t_0,t}}^{-1}=U_{t,t_0}
\end{align}
is implied immediately.
Accordingly, we have
\begin{align}\label{eq5.32}
  \brel\sigma(t)=U_{t,t_0}\brel\sigma(t_0)U_{t_0,t}\,,\qquad
  \brel\tau(t)=U_{t,t_0}\brel\tau(t_0)U_{t_0,t}\,,
\end{align}
and therefore 
$U\bigl(t_0;\brel\sigma(t),\brel\tau(t),t\bigr)= 
U\bigl(t_0;\brel\sigma(t_0),\brel\tau(t_0),t\bigr)$, that is to say: it
doesn't matter if we regard $U_{t,t_0}$ as a function of the spin operators
at the initial or the final time.

Upon combining (\ref{eq5.30}) with the Schr\"odinger equation
(\ref{eq5.23}), we find first
\begin{align}\label{eq5.33}
  i\hbar\frac{d}{dt}U_{t_0,t}=U_{t_0,t}H_t\,,
\end{align}
and then with Heisenberg's general equation of motion (\ref{eq5.29}),
\begin{align}\label{eq5.34}
i\hbar\frac{\partial}{\partial t}U_{t_0,t}=H_tU_{t_0,t}\,.
\end{align}
To begin with the implicit $\brel\sigma$ and $\brel\tau$ are all at time $t$,
but since the derivative in (\ref{eq5.34}) refers to the parametric
dependence on $t$, not to the dynamical one, we can just as well take all
$\brel\sigma$ and $\brel\tau$ at time $t_0$, so that the $t$ dependence is
then only parametric. 

With the understanding, then, that we are only dealing with parametric $t$
dependences, we can combine  (\ref{eq5.34}) with the initial condition
$U_{t_0,t_0}=\pmb{1}$ into the integral equation
\begin{align}\label{eq5.35}
 U_{t_0,t}=\pmb{1}-\frac{i}{\hbar}\int_{t_0}^tdt_1H_{t_1}U_{t_0,t_1}\,,   
\end{align}
which can be iterated to produce a formal series expansion, a variant of both
the Born series of scattering theory and the Dyson series of field theory, 
of $U_{t_0,t}$ in powers of the (time-integrated) Hamilton operator.
It begins with the terms
\begin{align}\label{eq5.36}
   U_{t_0,t}=\pmb{1}-\frac{i}{\hbar}\int_{t_0}^tdt_1H_{t_1}
   -\frac{1}{\hbar^2}\int_{t_0}^tdt_1\int_{t_0}^{t_1}dt_2H_{t_1}H_{t_2}
+\cdots\,.
\end{align}

For the Hamilton operator (\ref{eq5.22}), the Heisenberg equations of motion
(\ref{eq5.27}) and (\ref{eq5.28}) read  
\begin{equation}\label{eq5.37}
\left\{\begin{array}{l}
\dfrac{d}{dt} \brel\sigma(t) = 
\brel\Omega_1(t) \times\brel\sigma(t) - \omega(t) \brel\sigma(t) \times 
\brel\tau(t)\,,\\
\dfrac{d}{dt}\brel\tau(t) = 
\brel\Omega_2(t) \times \brel\tau(t) + \omega(t) \brel\sigma(t) 
\times\brel\tau(t)\,.\end{array}\right.
\end{equation}

\begin{exm}\label{exm5.1}
Let us give the solution of \eqref{eq5.34} for the case when 
$\brel\Omega_1(t)=0$,  
$\brel\Omega_2(t)=0$:
\begin{equation}\label{eq5.38}
\left\{\begin{array}{l}
  \brel\sigma(t) =\dfrac{1}{2}\bigl[\brel\sigma(t_0)+\brel\tau(t_0)\bigr]
           +\dfrac{1}{2}\bigl[\brel\sigma(t_0)-\brel\tau(t_0)\bigr]\cos\varphi
           -\dfrac{1}{2}\brel\sigma(t_0)\times\brel\tau(t_0)\sin\varphi
\,,\\
\noalign{\smallskip}
  \brel\tau(t) =\dfrac{1}{2}\bigl[\brel\sigma(t_0)+\brel\tau(t_0)\bigr]
           -\dfrac{1}{2}\bigl[\brel\sigma(t_0)-\brel\tau(t_0)\bigr]\cos\varphi
           +\dfrac{1}{2}\brel\sigma(t_0)\times\brel\tau(t_0)\sin\varphi,
           \end{array}\right.
\end{equation}
where
\begin{align}\label{eq5.39}
  \varphi=2\int_{t_0}^{t}dt'\omega(t')\,.
\end{align}
One can verify by inspection that (\ref{eq5.38}) solves (\ref{eq5.37}). 
$\hfill\square$
\end{exm}

Summarizing the above discussions, we have established the following.

\begin{thm}\label{thm5.7}
Let \eqref{eq5.3} and \eqref{eq5.4} be satisfied. Then we have
\begin{equation}
\left\{\begin{array}{l}
\dfrac{d}{dt} \pmb{\sigma}(t) = \dfrac1{i\hbar} [\pmb{\sigma}, H] (t) = 
\pmb{\Omega}_1(t) \times\pmb{\sigma}(t) - \omega(t) \pmb{\sigma}(t) \times 
\pmb{\tau}(t),\\
\noalign{\smallskip}
\dfrac{d}{dt}\pmb{\tau}(t) = \dfrac1{i\hbar} [\pmb{\tau}, H](t) = 
\pmb{\Omega}_2(t) \times \pmb{\tau}(t) + \omega(t) \pmb{\sigma}(t) 
\times\pmb{\tau}(t)
\end{array}\right.
\end{equation}
and 
\begin{equation}
\left\{\begin{array}{l}
\pmb{\sigma}(t) = U_{t,t_0}\pmb{\sigma}(t_0) U_{t_0,t}\\
\pmb{\tau}(t) = U_{t,t_0} \pmb{\tau}(t_0) U_{t_0,t}\end{array}\quad ,\quad 
\text{for 
any}\quad t_0,t\ge 0.\right.
\end{equation}
$\hfill\square$
\end{thm}

\section{Conclusion}\label{sec6}

New or improved versions of the elementary devices cited in this paper are 
constantly emerging. For example, in the design of cavity QED, 
instead of using a particle beam to shoot atoms through an optical 
cavity as illustrated in 
Fig.~3.2, there have been proposals to enclose ion traps inside a cavity. 
Thus, 
such proposals incorporate both features of ion traps and cavity QED as
studied  in this paper. 
Even though it is not possible for us to describe {\em all\/} 
major types of contemporary quantum computing devices or proposals, we hope 
that 
our mathematical analysis and derivations herein have provided reasonably 
satisfactory rigor and basic principles useful in helping future 
interdisciplinary research in quantum computation.

From the control-theoretic point of view, the best theoretical model for laser 
and quantum control  seems to be 
the Schr\"odinger equation offered by (\ref{eq5.3}) with the Hamiltonian given 
by (\ref{eq5.2}), wherein the $\pmb{B}_j(t)$'s and $J_{jk}(t)$'s
 are the {\em bilinear 
controllers\/}. 
Equivalently, the Heisenberg's equations of motion as presented in Theorem 
\ref{thm5.7} (which deal with only two quantum dots) form a {\em 
bi-trilinear control system\/} with $\pmb{\Omega}_1(t)$ and
 $\pmb{\Omega}_2(t)$ 
 therein as bilinear  controllers, while $\omega(t)$ is a {\em trilinear\/}
controller. 
Such control systems 
(as presented in Theorem \ref{thm5.7}, e.g.) look nearly identical to the 
classical rigid body rotation dynamics and, thus, a fair portion of the 
existing 
control results appear to be readily applicable. In our discussion in Section 
\ref{sec5}, we have derived certain formulas for the shaping of laser pulses 
(see (\ref{eq5.11}), (\ref{eq5.16}), (\ref{eq5.17}), \eqref{eq5.21a}, etc.) 
which can achieve the desired elementary 1-bit and 2-bit elementary quantum 
gates. However, we wish to emphasize that such laser pulse  shaping is {\em 
underutilized\/} as far as the control effects are concerned. To obtain more 
optimal control effects, {\em feedback strategies\/} need to be evaluated. 
Quantum feedback control is currently an important topic 
in quantum science and technology.

For the models of 2-level atoms, cavity QED and ion traps, 
our discussions here have shown that the admissible class 
of controllers is quite limited, the reason 
being that the control or excitation laser must operate at specified 
frequencies in order to stay within the two or three levels, 
and to avoid the excitation of unmodeled dynamics. 
We can exercise control mainly through the duration of the 
activation of laser pulses, such as \eqref{eq3.13a} has shown.

There are many interesting mathematical problems in the study of laser-driven 
control systems. Quantum computing/computer contains an abundant source 
of such 
problems for mathematicians and control theorists to explore.

\end{document}